\documentclass[a4paper,11pt,final]{article}

\pdfoutput=1 
\usepackage{jheppub} 
\usepackage{amssymb}
\usepackage{amsmath}
\usepackage{setspace}
\usepackage{amsfonts}
\usepackage{xcolor}
\usepackage{color}
\usepackage{braket}
\usepackage{bbm}
\usepackage{mathtools}
\usepackage{caption}
\usepackage{subcaption}
\usepackage{enumerate}
\usepackage{bm}
\usepackage{mathrsfs}
\usepackage{cleveref}
\usepackage[inline]{showlabels}
\usepackage[normalem]{ulem}
\usepackage{xparse}
\usepackage{bm}

\definecolor{dolphingreen}{rgb}{0,0.74,0.63}
\definecolor{dolphinorange}{rgb}{0.98,0.53,0.098}
\definecolor{purple}{rgb}{0.4,.2,0.7}
\usepackage{graphicx}
\graphicspath{{./Figs/}}

\usepackage{tikz}
\usepackage[table]{xcolor}
\usetikzlibrary{knots,hobby,calc,decorations.markings,decorations.pathmorphing,decorations.pathreplacing,fadings}

\DeclareMathOperator{\Tr}{Tr}
\DeclareMathOperator{\sgn}{sgn}

\newcommand{\tr}{\textrm{Tr}}

\newcommand{\cO}{\mathcal{O}}
\newcommand{\cA}{\mathcal{A}}

\def\bne{\begin{equation}}
	\def\ene{\end{equation}}
\def\beq{\begin{equation}}
        \def\eeq{\end{equation}}
\newcommand{\del}{\partial}

\newcommand{\dd}{\text{d}}
\newcommand{\pa}{\partial}
\newcommand{\mc}[1]{\mathcal{#1}}
\newcommand{\leftderiv}{\pa_v^{\scriptscriptstyle{(\downarrow)}}}
\newcommand{\rightderiv}{\pa_v^{\scriptscriptstyle{(\uparrow)}}}
\newcommand{\cUV}{c_{\scriptscriptstyle\mathrm{UV}}}
\newcommand{\cIR}{c_{\scriptscriptstyle\mathrm{IR}}}
\newcommand{\cT}{c_{\scriptscriptstyle\mathrm{T}}}

\DeclareMathOperator\erf{erf}

\makeatletter
\NewDocumentCommand{\boundName}{o}{%
  QNEI%
  \IfNoValueF{#1}{#1}%
  \futurelet\@let@token\@boundName@space
}
\newcommand{\@boundName@space}{%
  \ifx\@let@token.\else
    \ifx\@let@token,\else
      \ifx\@let@token;\else
        \ifx\@let@token:\else
          \ifx\@let@token!\else
            \ifx\@let@token?\else
              \ifx\@let@token)\else
                \space
              \fi
            \fi
          \fi
        \fi
      \fi
    \fi
  \fi
  \ignorespaces
}
\makeatother

\newsavebox{\imagebox}

\title{

Curious QNEIs from QNEC: New Bounds on Null Energy in Quantum Field Theory

}
\date{}
\author[a,b]{Jackson R. Fliss,}
\author[c]{and Andrew Rolph}
\affiliation[a]{Department of Applied Mathematics and Theoretical Physics, University of Cambridge, Cambridge CB3 0WA, UK}
\affiliation[b]{Physique Th\'eoretique et Math\'ematique, Universit\'e Libre de Bruxelles \& International Solvay Institutes, CP 231, 1050 Bruxelles, Belgique}
\affiliation[c]{Vrije Universiteit Brussel (VUB) and The International Solvay Institutes, Pleinlaan 2, B-1050 Brussels, Belgium}

\abstract{
We derive new families of quantum null energy inequalities (QNEIs), i.e. bounds on integrated null energy, in quantum field theories in two and higher dimensions. 
These are universal, state-independent lower bounds on semi-local integrals of $\langle T_{vv} \rangle$, the energy-momentum flux in a null direction,
and the first of this kind for interacting theories in higher dimensions. Our ingredients include the quantum null energy condition (QNEC), strong subadditivity of von Neumann entropies, defect operator expansions, and the vacuum modular Hamiltonians of null intervals and strips. These results are fundamental constraints on null energy in quantum field theories.}

\begin{document}

\maketitle

\section{Introduction}\label{sect:intro}

Energy plays a ubiquitous role in physics; it underlies dynamics in all systems, connects matter to geometry in gravity, and is central to thermodynamics and statistical physics.
Many physical classical field theories obey pointwise energy conditions, and these have played an important role in, for example, singularity theorems~\cite{penrose1965gravitational}.
However, for local, relativistic quantum field theories (QFTs), the study of energy is both richer and more precarious.
Due to the short-distance entanglement of the vacuum, the energy-momentum density at any point can be made arbitrarily negative~\cite{epstein1965nonpositivity}.

To get a lower bound on the stress tensor operator in QFTs, it must be `smeared' -- integrated against a test function -- over a spacetime domain.
Such constraints are called {\it quantum energy inequalities} (QEIs)~\cite{Fewster:2012yh}.
In general, a QEI can express the lower bound on the energy density in terms of the expectation values of other operators; however, particularly powerful are {\it state-independent} QEIs which express smeared energy densities in terms of a state-independent functional of the smearing, schematically:
\beq\label{eq:SIQEIschem}
    \int_D\dd^px\,g(x)\langle \xi_1^\mu \xi_2^\nu T_{\mu\nu}\rangle_\psi\geq - \mc F_{\xi_{1,2}}[g]~.
\eeq
$g$ is a smooth and positive smearing function with support on a spacetime region $D$. The dimension $p$ of $D$ varies between QEIs; anything from a one-dimensional curve to a codim-0 subregion.
$\mc F_{\xi_{1,2}}$ is a functional of $g$ that depends on the components of the stress-energy tensor being considered,
as well as other details of the quantum field theory (e.g. correlation lengths). 

From here onwards, whenever we use the term `QEI', we will mean a state-independent bound of the form \eqref{eq:SIQEIschem}.
Perhaps the best known QEI is the {\it averaged null energy condition} (ANEC), 
which states that the null stress tensor integrated uniformly over an entire (achronal) null geodesic is positive in all states:
\beq
\qquad\qquad\qquad\qquad   \int_{-\infty}^\infty \dd v\,\langle T_{vv}\rangle\geq 0~,\qquad\qquad\qquad \text{(ANEC)}
\eeq
where $v^{\mu}$ is the null vector generating the geodesic. 
Among other applications, achronal ANEC is necessary for causality in AdS/CFT~\cite{Gao:2000ga}, and forbids long traversable wormholes~\cite{morris1988wormholes, Hochberg:1998ii}.
However, of additional interest are QEIs of the more semi-local variety, where, unlike in ANEC, $D$ is a compact domain. 
Due to their state-independence and semi-locality, such QEIs have been a key tool in extending the classical singularity theorems of general relativity (such as those by Hawking and Penrose~\cite{penrose1965gravitational, Hawking:1966sx, Hawking:1966jv, Hawking:1967ju}) to semi-classical gravity coupled to QFT~\cite{Brown:2018hym, Fewster:2019bjg, Fewster:2021mmz}. 

While the ANEC has been shown to hold for all QFTs in Minkowski spacetime~\cite{Faulkner:2016mzt, Hartman:2016lgu}, and for free theories in certain classes of curved spacetimes~\cite{Wald:1991xn, Kelly:2014mra, Kontou:2015yha}, deriving and proving QEIs of the semi-local variety is a difficult endeavour. Perhaps the best known example is Fewster and Hollands' proof that all two-dimensional conformal field theories (CFTs) admit the following semi-local QEI on the stress tensor \cite{Fewster:2004nj}
\beq\label{eq:FHbound}
   \qquad\qquad\qquad\qquad\qquad\int\dd v\,g(v)\langle T_{vv}\rangle\geq -\frac{c}{48\pi}\int\dd v\frac{g'(v)^2}{g(v)}~,\qquad\text{(FH bound)}
\eeq
where $c$ is the central charge of the CFT. See also~\cite{Blanco:2017akw} for a stronger but state-dependent version of~\eqref{eq:FHbound}, and~\cite{Blanco:2013lea}, which derives bounds on how far negative energy can be isolated away from positive energy. Both of those papers use monotonicity of relative entropy, which we will use too. 
The first result in this paper will be a new, elementary proof and extension of~\eqref{eq:FHbound}.

A different extension of the FH bound to states of $d>2$ QFTs consistently coupled to semi-classical gravity, coined the {\it smeared null energy condition} (SNEC), was conjectured by Freivogel and Krommydas \cite{Freivogel:2018gxj} 
\beq\label{eq:snecdef}
    \qquad\qquad\qquad\qquad\int\dd v\,g(v)\langle T_{vv}\rangle\geq-\frac{1}{32\pi G_N}\int\dd v\frac{g'(v)^2}{g(v)}~,\qquad\text{(SNEC)}
\eeq
in which the central charge is replaced by the inverse Newton's constant. The key differences with respect to the FH bound are the coupling to gravity and that $d>2$. The SNEC has been proven for holographic CFTs coupled to induced gravity in a braneworld model \cite{Leichenauer:2018tnq}, however, broader evidence for the SNEC has so far remained elusive. This is partially because, as a gravitational bound, gravitational physics must be part of its derivation.

Amongst the QEIs that can be established purely in QFTs in greater than two dimensions, almost all results have been established for free QFTs (with ANEC the notable counterexample). 
Furthermore, it is known that certain free theories do \textit{not} admit state-independent lower bounds: an example is the non-minimally coupled scalar field for which one can construct states of arbitrarily negative expectation value of the timelike or null-contracted stress tensor, regardless of the domain over which it is smeared \cite{Fewster:2007ec,Fliss:2023rzi}. This 
non-existence
of a QEI is pertinent to the free scalar CFT even in flat space, where the non-minimal coupling manifests as an improvement term needed to make the stress tensor traceless. 

It is natural to wonder whether this QEI non-existence is particular to free field theories, or whether QEIs are generically too much to hope for. 
Before this paper, this question remained largely unanswered, though investigations into large-$N$ CFTs with a `generalized free field' structure have indicated that interactions may prevent the `pile-up' of negative energy density allowed in free theories \cite{Fliss:2024dxe}. 

We will derive new QEIs on null energy, and so answer this question.
To do so, we will make use of a completely complementary approach from those used so far in deriving QEIs, one that makes use of the connection between energy and entropy,
through the 
{\it quantum null energy condition} (QNEC) \cite{Bousso:2015mna}, which bounds the local expectation value of the null stress tensor by a second-order shape derivative 
of the entanglement entropy at that point:
\beq\label{eq:QNECschem}
   \qquad\qquad\qquad\qquad \langle T_{vv}(v)\rangle\geq\frac{1}{2\pi}\pa_v^2S~.\qquad\qquad\qquad \text{(QNEC)}
\eeq
The above formula is schematic; in the body of this paper, we will provide more precise details.
The QNEC lends a universality to our results, because it has been proven for free and super-renormalisable~\cite{Bousso:2015wca}, holographic~\cite{Koeller:2015qmn}, and generic interacting QFTs \cite{Balakrishnan:2017bjg, Ceyhan:2018zfg, Hollands:2025glm}. 
Moreover, for $d>2$ interacting theories, there is strong evidence that this local form of the QNEC is saturated~\cite{Leichenauer:2018obf, Balakrishnan:2019gxl}, implying that it is the tightest possible bound on local energy densities. Note that ANEC follows from integrating QNEC. 
The entanglement entropy that appears in~\eqref{eq:QNECschem} obeys its own inequalities -- weak monotonicity, strong subadditivity (SSA), and others -- and we will leverage these in our derivations to go from entropy-dependent to state-independent bounds. 

In the gravitational context, QNEC, and the {\it quantum focussing conjecture} (QFC) from which it descends~\cite{Bousso:2015mna}, connect energy and entropy, which has been insightful for holography and semi-classical gravity~\cite{Faulkner:2022mlp}.
Related inequalities, such as the generalised second law \cite{Bekenstein:1974ax} and the Bousso bound \cite{Bousso:1999xy, Rolph:2022csa}, are also effective starting points for constraining semi-classical gravity in a manner complementary to QEI-based approaches~\cite{Wall:2010jtc,Bousso:2022cun,Bousso:2022tdb,Bousso:2025xyc}.

However, from a purely field-theoretic, non-gravitational point of view, it is not clear that the QNEC, even when it is saturated, is an effective bound on the energy density. 
For one, there are states for which the right-hand side of \eqref{eq:QNECschem} 
is arbitrarily negative
(consistent with the pointwise stress tensor being an unbounded operator). 
Secondly, at a practical level, the evaluation of entanglement entropy is extremely difficult, except for certain symmetric regions and for holographic theories~\cite{Ryu:2006bv}.
Lastly, there is an undesirable mismatch between the linear (on the left-hand side) and non-linear (on the right-hand side) state dependences of \eqref{eq:QNECschem}. 

In this paper, we will derive new state-independent bounds on smeared null energy in generic interacting QFTs. We call these \textit{quantum null energy inequalities} (\boundName[s]).
Our method will be to start by integrating QNEC against a smearing function, and then, to make the bound state-independent, use entanglement entropy inequalities, vacuum modular Hamiltonians, and defect operator product expansions (OPEs). 

Firstly, in Section \ref{sec:2dS}, we show that the `conformally improved' QNEC~\cite{Wall:2011kb} directly implies the FH bound \eqref{eq:FHbound} and extends it to all super-renormalizable QFTs. Our derivation is extremely simple, yet, as far as we are aware, has not yet been presented in the literature. 

Subsequently, in Section \ref{sect:Rolphbound}, we show how the QNEC, along with monotonicity of relative entropy, imply an infinite family of \boundName[s] in generic 2d QFTs parametrised by a function $\zeta$. To be specific, given a positive and localised smearing function $g$, our bounds take the following form
\beq\label{eq:intro2dresult}
    \boxed{\int\dd v\, m(v)\langle T_{vv}\rangle\geq-\frac{\cUV}{12\pi}\int\dd v\frac{g'(v)}{v-\zeta(v)}~,}
\eeq
where $\cUV$ is the central charge of the UV-fixed point of the theory and $m(v)$ is a convolution of $g$ with a kernel associated to the modular Hamiltonian of a null strip whose precise form is given in \eqref{eq:mdef}. Much like $g$ itself, $m$ is positive and localised (e.g. has compact support when $g$ has compact support). 
The function $\zeta$ essentially parametrizes the choice of null strips as entangling regions in our monotonicity argument and appears in the definition of $m$. 

A simple grounding example is given by choosing $\zeta(v) =v+ \sgn(v) \delta$, with $\delta >0$. Then the right hand side of~\eqref{eq:intro2dresult} simply becomes $-\cUV g(0)/(6\pi \delta)$. The relevant $m$ to which it applies is also straightforwardly computable depending on the initial smearing function. We can, e.g., let $g$ be a Gaussian. Then the correction to the original smearing function, $h= m-g$, is given by~\eqref{eq:hvhva} and plotted in Fig.~\ref{fig:hvplot}.

In Section \ref{sect:highd}, we adapt our methods to interacting CFTs in $d>2$ dimensions. By considering the monotonicity argument applied to `nearly null' strips, we derive an infinite family of \boundName[s] smeared over a causal diamond uniform in $d-2$ transverse directions. Again, to be specific, given a positive and localized function $g(v)$ of the lightcone coordinates, we derive
\beq\label{eq:introhighdresult}
    \boxed{\begin{split}\int\dd^{d-2}y_\perp&\int\dd v\mc M (v)\langle T_{vv}\rangle\\
    &\geq-\frac{\beta\,\cT}{2\pi}\int\dd^{d-2}y_\perp\int\dd v\left(\frac{g'(v)}{\epsilon_u(v)^{\frac{d-2}{2}}(v-\zeta(v))^{\frac{d}{2}}}+O(\epsilon_u)\right)~,
    \end{split}}
\eeq
where $\cT$ is the stress tensor two-point coefficient, and $\beta$ is a theory-dependent constant that we will define later. $\mc M$ is defined completely analogously as $m$ in our $d=2$ bound: it is a particular convolution of $g$ with a kernel associated to the modular Hamiltonian of a null strip and depends on the choice of $\zeta$. Its precise definition is given in \eqref{eq:Mdef}. The new ingredient in \eqref{eq:introhighdresult} in comparison to \eqref{eq:intro2dresult} is the additional function $\epsilon_u$, and that \eqref{eq:introhighdresult} applies when $\epsilon_u\ll1$. The origin of $\epsilon_u$ is a parameterization of how far away the `nearly null' strip regions (that appear at an intermediate step in our derivation) are from being truly null. 
The leading-order, non-positive-power terms in~\eqref{eq:introhighdresult} are zero or state-independent, but the linear order $\epsilon_u$ terms are state-dependent; 
our bound is approximately state-independent for small $\epsilon_u$.
While the uniform transverse smearing imposes some limitations on the utility of our results (essentially introducing a divergent IR volume to the bound),
to our knowledge, our results present both the first instances of QEIs derived taking the QNEC as an input, and the first QEIs in a large class of interacting theories in greater than two dimensions. We see this as a major step in addressing the existence of state-independent bounds on locally smeared energy density in generic interacting theories.

\subsection*{Notational conventions}

We will denote the vacuum state of the QFT by $|\Omega\rangle$ and a generic state by $|\psi\rangle$. Expectation values will be denoted by a subscript, i.e.
\beq
    \langle \mc O\rangle_\psi=\langle\psi|\mc O|\psi\rangle~.
\eeq
When an expectation value is for a generic state, we will often omit the subscript. We will always denote vacuum quantities as such, however.

Entropies depend both on a region and a state. We will denote state dependence with a subscript and the region dependence with parentheses. The complement of a region, $A$, will be denoted as $A^c$. Much like expectation values, when the context is clear for a generic state, we will often omit this subscript. Vacuum entropies will always be denoted as such with a subscript. To give an example of our notation, let us recall the definition of the vacuum entanglement entropy here:
\beq    \rho_\Omega(A)=\tr_{A^c}\ket{\Omega}\bra{\Omega}~,\qquad S_\Omega(A)=-\tr_{A}\left(\rho_\Omega(A)\log\rho_\Omega(A)\right)~.
\eeq
We will work in $d$-dimensional Minkowski spacetime in `mostly plus' signature:
\beq
    \dd s^2=\eta_{\mu\nu}\dd x^\mu \dd x^\nu=-\dd t^2+\delta_{ij}\dd x^i\dd x^j~.
\eeq
We will use lightcone coordinates $v=t+x^1$ and $u=t-x^1$, denoting the remaining coordinates on $\mathbb R^{d-2}$ as $\vec y_\perp$:
\beq
    \dd s^2=-\dd u\dd v+\sum_{i=2}^d\dd y_\perp^i\dd y_\perp^i~.
\eeq
	
\section{Two-dimensional quantum null energy inequalities}\label{sec:2dSNEC}
	
\subsection{Fewster-Hollands bound from 2d QNEC}\label{sec:2dS}
Consider a QFT on 2d Minkowski spacetime and let $T_{vv} (v)$ be the $vv$ component of its stress tensor at the point $p = (0,v)$. The choice of $u = 0$ is without loss of generality.
The QNEC for 2d QFTs is~\cite{Wall:2011kb, Bousso:2015mna, Koeller:2015qmn}%
\bne \label{eq:2dQNE}
    2\pi \langle T_{vv} (v) \rangle \geq \del_v^2 S(v) + \frac{6}{\cUV}(\del_v S(v))^2 ~,
\ene
where $\cUV$ is the central charge of the CFT at the QFT's UV fixed point.  $S(v)$ is the entanglement entropy of the state $\psi$ in question reduced on any achronal curve with one endpoint at $p$, such as a semi-infinite curve ending at the $(0,v)$ and extending to infinity%
\footnote{If the global state is pure, then $S(v)$ is the same whether the half-line extends to the left or right. If the state is mixed, then this is no longer true, and there are separate QNEC inequalities for the left and right half-lines.}.

The 2d QNEC~\eqref{eq:2dQNE} has been proven for 2d CFTs and relevant deformations of holographic 2d CFTs~\cite{Koeller:2015qmn}, but not in complete generality. 
Compared to QNEC in higher dimensions, the 2d QNEC is special because it has the $(\del_v S)^2$ term, which is absent from the higher-dimensional QNEC. In CFTs, the 2d QNEC is saturated by Ba\~nados states (Virasoro coadjoint orbit states)~\cite{Khandker:2018xls, Ecker:2019ocp}. The $(\pa_vS)^2$ term makes the right-hand side of~\eqref{eq:2dQNE} transform as a conformal primary, which means that both sides of the inequality have the same conformal anomaly, making it independent of the conformal frame~\cite{Wall:2011kb}. 

Let $g(v)$ be a smooth, non-negative, smearing function which is either a bump function or a Schwartz function.%
\footnote{Bump functions, also known as test functions, are smooth and have compact support. An example would be $g(v) = \exp((v^2-\sigma^2)^{-1})$ for $|v|\leq \sigma$ and $0$ for $|v| > 0$. Schwartz functions, such as the Gaussian function, decay faster than any power at infinity.}
We integrate both sides of~\eqref{eq:2dQNE} against $g(v)$, and then integrate the 
$\del_v^2 S$ term
by parts, giving
\bne
    2\pi \int \dd v g(v) \langle T_{vv} \rangle \geq  \int \dd v \left(\frac{6}{\cUV}g(v)(\del_v S)^2- g'(v) \del_v S \right ) + \left[g(v) \del_v S \right]^\infty_{-\infty} ~.\label{eq:intSn} 
\ene
We assume that the boundary term vanishes, which is true if $g(v)$ is a bump function, and a mild assumption on $\del_v S$ at infinity if $g(v)$ is a Schwartz function.
Next, we take the rest of the right-hand side of~\eqref{eq:intSn}, complete the square for the $\del_v S$ terms, and drop the resulting non-negative, squared term from the inequality, and this gives
\bne
    \boxed{\int \dd v g(v) \langle T_{vv} \rangle \geq - \frac{\cUV}{48\pi} \int \dd v \frac{g'(v)^2}{g(v)}~. }\label{eq:bound} 
\ene
This is precisely the Fewster-Hollands (FH) bound for 2d CFTs \cite{Fewster:2004nj} introduced in Section \ref{sect:intro}. 

To give an example and gain some intuition for the
FH bound, let us take $g(v)$ to be a normalised Gaussian of width $\sigma$, for which~\eqref{eq:bound} becomes
\bne
    \int \dd v g(v) \langle T_{vv} \rangle \geq - \frac{\cUV}{48\pi\sigma^2}~. \label{eq:fhgau}
\ene
If we take $\sigma \to \infty$, then this bound reduces to ANEC, and the bound becomes trivial in the $\sigma \to 0$ limit. The latter limit is in accordance with the general expectation that pointwise energy densities are unbounded in QFT.

Some other comments on the 
FH bound~\eqref{eq:bound}:
\begin{itemize}
    \item The right-hand side is state-independent, and also non-positive definite, as it must be for a state-independent energy bound given that the LHS is zero for the vacuum state.
    \item Unlike the 2d QNEC, ~\eqref{eq:2dQNE}, the FH  bound is not saturated for the vacuum state for the simple reason that $\langle T_{vv}\rangle_\Omega=0$. However, at the UV fixed point, the bound is tight and the states saturating it can easily be constructed \cite{jrf_modave} as
    \beq
        |\psi\rangle=U[g]|\Omega\rangle~,
    \eeq
    where $U[g]$ is the unitary operator implementing the conformal transformation
    \beq
        U[g]T_{vv}(v)U[g]^\dagger = (\tilde v'(v))^2T_{\tilde v\tilde v}(\tilde v(v))-\frac{\cUV}{24\pi}\{\tilde v,v\}~,
    \eeq
    with $\{\cdot,\cdot\}$ and Schwarzian derivative and $\tilde v(v)$ chosen such that $\tilde v'(v)=1/g(v)$.
    \item The FH 
    bound \eqref{eq:bound} only holds in two dimensions; there are explicit counterexamples in higher dimensions. Namely, it was shown by Fewster and Roman that the left-hand side of \eqref{eq:bound} is unbounded in the class of ``0+2 states" (superpositions of zero and two-particle states) of free scalar field theory in $d=4$ dimensions \cite{Fewster:2002ne}. More generally, one can construct a series of squeezed states in free scalar field theories of arbitrary dimension $d>2$ where the left-hand side of~\eqref{eq:bound} can be made arbitrarily negative \cite{Fliss:2021gdz}. However, when the free theory comes equipped with a UV cutoff on the transverse momentum, then the properties of free theories quantized on lightsheets \cite{Burkardt:1995ct,Wall:2011hj} allow one to ``bootstrap'' the FH 
    bound to higher dimensions:
    \beq
        \int \dd v\,g(v) \langle T_{vv}(v)\rangle \geq -\frac{\cUV}{48\pi\,a^{d-2}}\int\dd v\frac{(g')^2}{g}~.
    \eeq
    Here $a$ is the `pencil width' of the lightsheet discretized in the transverse directions, which provides a natural (inverse) UV cutoff on the transverse momenta, and $\cUV$ is simply the number of independent scalar fields.

    \item Our derivation of~\eqref{eq:bound} is novel because it is both elementary and extends the proof of this FH bound to all 2d QFTs for which~\eqref{eq:2dQNE} is proven, which includes some non-conformal theories, namely relevant deformations of holographic 2d CFTs~\cite{Koeller:2015qmn}. 

    \item Lastly, as mentioned above, the coefficient $\cUV$ appearing in \eqref{eq:bound} is the central charge of the QFT's UV fixed point. However, at the other end of the RG flow, the QFT is described by an IR CFT with central charge $\cIR$. The stress tensor of this theory will satisfy the FH 
    bound with coefficient $\cIR$, which can be proven directly in the CFT. This is consistent with our claim due to the monotonicity of the central charge under RG flows \cite{Zamolodchikov:1986gt},
    \beq
        \cIR \leq \cUV~.
    \eeq
\end{itemize}

\subsection{An family of 2d QNEIs}\label{sect:Rolphbound}
In the interest of investigating the utility of the QNEC for establishing QNEIs in higher dimensions, we will look at deriving a version of the 
FH bound that does not make use of the $(\pa_v S)^2$ term of \eqref{eq:2dQNE} since this term doesn't appear in $d>2$. If it did, then, using the same method as in section~\ref{sec:2dS}, we would have a state-independent lower bound on $\int dv g(v) \langle T_{vv} \rangle$ in $d>2$, which is impossible~\cite{Fewster:2002ne}. 
We will work exclusively in $d=2$ QFTs in this subsection, leaving an exploration to higher-dimensional QFTs in Section \ref{sect:highd}. We will find that this will lead to an infinite family of QEIs, parameterized by a function $\zeta$, which are interesting in their own right. 

To begin, we simply drop the $(\pa_vS)^2$ term from \eqref{eq:2dQNE} and start from the weaker inequality
\bne 
    2\pi\langle T_{vv} \rangle \geq \del_v^2 S \label{eq:2dQN2} . 
\ene
So, with respect to the FH 2d 
bound, the \boundName[s] we derive will be proven for any 2d QFT for which~\eqref{eq:2dQN2} is proven, which includes super-renormalisable QFTs~\cite{Bousso:2014sda}.

\subsubsection{Integrating QNEC}
Integrating~\eqref{eq:2dQN2} against $g(v)$, and then integrating by parts gives
\bne \label{eq:integratedQNEC}
    \int_{-\infty}^{\infty} \dd v g(v) \langle T_{vv}  (v) \rangle  \geq - \frac{1}{2\pi}\int_{-\infty}^{\infty} \dd v~g'(v)~\del_v S(v)~.
\ene

Generally, $g'(v)$ in~\eqref{eq:integratedQNEC} will take both signs, and so, to lower bound the RHS of~\eqref{eq:integratedQNEC} in an $S$-independent way, we will need both lower and upper bounds on $\del_v S(v)$. For simplicity, though this can easily be generalised, we will assume that $g'(v)$ switches sign only once, which can be taken, without loss of generality, to be at $v=0$, i.e.
\beq
    -\text{sgn} (v) g'(v) \geq 0~.
\eeq
A canonical example of which to keep in mind would be the normalized Gaussian,
\beq\label{eq:normGauss}
    g(v) = \frac{e^{-\frac{v^2}{2\sigma^2}}}{\sqrt{2\pi\sigma^2}}~.
\eeq 

\subsubsection{Bounds on first derivatives of entropy}

We will derive bounds between $\del_v S(v)$ using strong subadditivity and monotonicity of relative entropy.%
\footnote{As an aside, note that SSA and monotonicity of relative entropy are equivalent~\cite{lieb1973proof}, which suggests there exists a possible simplification of our following construction.}
The bound will be in terms of the $\del_v$ derivative of vacuum-subtracted modular Hamiltonians and vacuum entropies. 
	
\paragraph{Upper bound.} First we derive an upper bound on $\del_v S(v)$. 
Strong subadditivity states that for three regions, $X$, $Y$, and $Z$,
\bne 
    S(XYZ) - S(XY) \leq S(YZ) - S(Y)~. \label{eq:SSAv1}
\ene

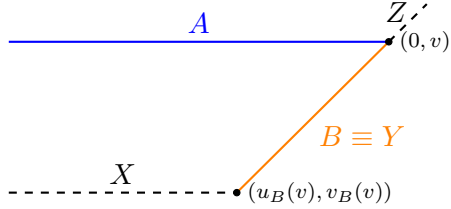
\begin{figure}[h!]
    \centering
    \begin{tikzpicture}
        \draw[thick,blue] (-5,0) to (0,0) node[right, black] {\scriptsize $(0,v)$};
        \draw[thick,orange] (-2,-2) to (0,0);
        \filldraw[black] (-2,-2) circle (1.2pt);
        \filldraw[black] (0,0) circle (1.2pt);
        \draw[thick,dashed] (-5,-2) to (-2,-2) node[right, black]{\scriptsize $(u_B (v),v_B(v))$};
        \draw[thick,dashed] (0,0) to (.5,.5);
        \node[blue] at (-2.5,.25) {$A$};
        \node[orange] at (-.35,-1.25) {$B \equiv Y$};
        \node at (-3.5,-1.75) {$X$};
        \node at (0.1,.4) {$Z$};
    \end{tikzpicture}
    \caption{The regions $X$, $Y$, $Z$, $A$ and $B$ as used in~\eqref{eq:SSAv1} to prove~\eqref{eq:upperbound}. }
\label{fig:upper_bound}    
\end{figure}

Take $X$ to be a semi-infinite half-line extending to the left, $Y$ to be any interval adjoining $X$ on the right, and $Z$ to be an infinitesimal null interval that extends $XY$ further in the positive $v$ direction. See Figure~\ref{fig:upper_bound}. 

In Figure~\ref{fig:upper_bound}, we have depicted $Y$ as a null interval, because that is the special case that we will be interested in soon. $\del_v$ is null, future-directed, and outward with respect to $XY$, i.e. to the right. 

Next, choose any region $A$ for which $D(A) = D(XY)$ and $B$ any region whose domain of dependence satisfies $D(B) \subseteq D(A)$, and for which $\del A \cap \del B$ includes the point at which the derivative is taken. Above, we have taken $B=Y$. Without loss of generality, we will take the endpoint of $A (v)$ to be at $(0,v)$. Then $B(v)$ is any achronal curve between the endpoints at $(u_B(v),v_B(v))$ and $(0,v)$.  We will denote this as
\bne 
    B(v) = [(u_B(v),v_B(v)),(0,v)] \label{eq:Bdefn}~.
\ene
Achronality requires $u_B (v) \geq 0$ and $v_B (v) \leq v$.

Then,~\eqref{eq:SSAv1} becomes
\bne \label{eq:upperbound} 
    \del_v S(A) \leq \, \rightderiv S(B)~,
\ene
where $\rightderiv$ means that the derivative acts on the upper boundary of $B$ shared with $A$. To wit, if 
$B' = [ (u'_1,v'_1) , (u'_2 , v'_2)]$, then
\bne 
    \rightderiv S (B(v)) := \del_{v'_2} S(B')\big|_{B' = B(v)}~.
    \label{eq:righd}
\ene

Next, we will bound $\rightderiv S(B)$. Note that, since QNEC applies to any achronal curve with an endpoint at $(0,v)$, including $B(v)$, we could chosen $B(v)$ when integrating QNEC, instead of the half-line $A(v)$, and then skipped straight to this step of bounding $\rightderiv S(B)$. But, because of~\eqref{eq:upperbound}, which followed from SSA alone, and the fact that the von Neumann entropies (in particular, the regions their reduced density matrices are defined for) will not be in the energy bounds we derive, the end result will be the same. We started from the half-line $A$ to emphasise this point.

We rewrite~\eqref{eq:upperbound} in terms of the vacuum-subtracted modular energy and the vacuum entropy. Starting from~\eqref{eq:upperbound}, we get
\bne \begin{split}
    - \del_v S(A) &\geq -\rightderiv S (B) \\
    &= -\rightderiv \Delta S(B) - \rightderiv S_{\Omega}(B) \\
    &= -\rightderiv(\Delta K_B - S_\text{rel}(\rho_B ||\rho_\Omega)) - \rightderiv S_{\Omega}(B)\\
    &\geq -\rightderiv\Delta K_B - \rightderiv S_{\Omega}(B)~. \label{eq:inmre}
\end{split} \ene 
This is the first inequality that we want. $\Delta K_B$ is the vacuum-subtracted expectation value of the vacuum modular Hamiltonian on $B$, i.e.

\bne
    \Delta K_B = - \Tr\left[(\rho(B) - \rho_{\Omega}(B))\log \rho_{\Omega}(B)\right]~,
\ene
where $\rho(B) = \Tr_{B^c} \rho$ is the reduced density matrix on $B$. 
In the last line of~\eqref{eq:inmre} we used monotonicity of relative entropy, and used that relative entropy can be written as
\bne
    S_\text{rel}(\rho||\sigma) = (\langle K_\sigma \rangle_\rho - \langle K_{\sigma} \rangle_\sigma) - (S_\rho - S_\sigma)~.
\ene
	
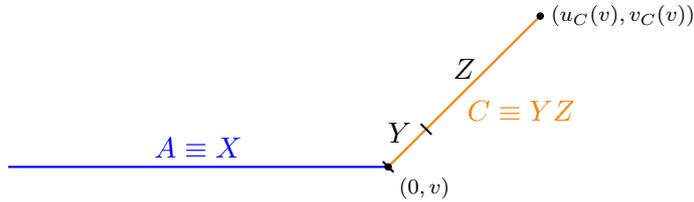
\begin{figure}[h!]
\centering
\begin{tikzpicture}
    \draw[thick,blue] (-5,0) to (0,0);
    \draw[thick,orange] (0,0) node[below right, black]{\scriptsize $(0,v)$} to (2,2) node[right, black]{\scriptsize $(u_C (v),v_C(v))$};
    \filldraw[black] (0,0) circle (1.2pt);
    \filldraw[black] (2,2) circle (1.2pt);
    \begin{scope}[shift={(.5,.5)},rotate=-45]
        \draw[thick] (-.1,0) to (.1,0);
    \end{scope}
    \begin{scope}[rotate=-45]
        \draw[thick] (-.1,0) to (.1,0);
    \end{scope}
    \node[blue] at (-2.5,.25) {$A\equiv X$};
    \node[orange] at (1.75,.75) {$C\equiv YZ$};
    \node at (1,1.3) {$Z$};
    \node at (.15,.45) {$Y$};
\end{tikzpicture}
\caption{The regions $X$, $Y$, $Z$, $A$ and $C$ as used in~\eqref{eq:lowerbound} to prove~\eqref{eq:sacbo}.\label{fig:AC}}
\end{figure}

\paragraph{Lower bound.} Next, we derive a lower bound on $\del_v S(v)$ using strong subadditivity in its second form, also known as weak monotonicity, which states that for regions $X$, $Y$, and $Z$,
\bne \label{eq:lowerbound}
    S(XY) - S(X) \geq S(Z) - S(YZ)~.
\ene
Take $X$ to be the same semi-infinite half-line extending to the left, but now $Y$ is an infinitesimal null interval adjoining $X$ in the direction of $\del_v$, and $Z$ is an arbitrary interval extending $XY$ further to the right. See Figure~\ref{fig:AC}. 

With respect to Figure~\ref{fig:upper_bound}, we have essentially switched $Y$ and $Z$. Next, relabel $X$ to $A$, and take any $C$ for which $D(C) = D(YZ) \subseteq D(A^c)$. Again, take $\del A (v) = (0,v)$, and then $D(C) \subseteq D(A^c)$ implies that $C (v)$ is
\bne 
    C(v) = [(0,v), (u_C(v), v_C(v))] ~,\label{eq:Cdefn}
\ene
with $u_C(v) \leq 0$ and $v_C (v) \geq v$. With these definitions,~\eqref{eq:lowerbound} becomes%
\footnote{The sign of $ \leftderiv S(C)$ in~\eqref{eq:sacbo} is a little counterintuitive because increasing $v$ decreases the size of $C$.}
\bne 
    \del_v S(A) \geq \, \leftderiv S(C) \label{eq:sacbo}~. 
\ene
$\leftderiv$ acts on the left boundary point of $C$, the shared boundary $\del A (v) \cap \del C (v) = (0,v)$, i.e. if $C' = [ (u'_1,v'_1) , (u'_2 , v'_2)]$, then%
\bne \leftderiv S(C) := \del_{v_1'} S(C')|_{C' = C(v)}. \label{eq:leftd}\ene
Again, we relate~\eqref{eq:sacbo} to vacuum-subtracted modular energy and vacuum entropy using monotonicity of relative entropy. Starting from~\eqref{eq:sacbo}, we get
\bne \begin{split}
    \del_v S(A) &\geq \,\leftderiv S(C) \\
    &= \, \leftderiv(\Delta K_C - S_\text{rel}(\rho_C ||\rho_\Omega)) +  \leftderiv S_{\Omega}(C) \\
    &\geq \, \leftderiv \Delta K_C +  \leftderiv S_{\Omega}(C)~.
\end{split} \ene
The region $C$ decreases with increasing $v$, the opposite of $B$, hence the opposite sign on the bound of the relative entropy derivative.

\paragraph{Upper and lower bound together.}
Bringing everything together, we have the following two-sided bound on $\del_v S(v) = \del_v S(A(v))$:
\bne 
    \boxed{ \leftderiv \Delta K_C + \leftderiv S_{\Omega}(C) \leq \del_v S(A) \leq \rightderiv \Delta K_B + \rightderiv S_{\Omega}(B)}~, \label{eq:magtb} 
\ene
where, recall, $B(v)$ and $C(v)$ are arbitrary intervals, except that each shares a boundary with $A(v)$, $D(B) \subset D(A)$, and $D(C) \subseteq D(A^c)$. The modular Hamiltonian and the vacuum entanglement entropy for generic 2d QFTs and arbitrary intervals are not known. However, for a given 2d QFT, if we assume that the null interval limit is controlled by the UV fixed-point CFT of the QFT, then we can use the known CFT results for $\Delta K$ and $S_{\Omega}$~\cite{Holzhey:1994we, Casini:2011kv}. We will utilise this in what follows. To take the null interval limit of $B(v)$ and $C(v)$, we set $u_B (v) \to 0$ in~\eqref{eq:Bdefn}, and $u_C(v) \to 0$ in~\eqref{eq:Cdefn}.
	
\subsubsection{Modular energies of null intervals}
Here we will derive the vacuum-subtracted modular energy for the interval $u=0, v \in [a,b]$ in 2d-Minkowski space, which is%
\footnote{Besides the stress tensor component, the vacuum modular Hamiltonian has a c-number part which drops out in the vacuum subtraction.}
\bne \label{eq:mod}
    \Delta K = 2\pi \int_a^b \dd v' \frac{(b-v')(v'-a)}{b-a} \langle T_{vv} (v') \rangle~.
\ene
This can be derived by taking the formula for the vacuum modular Hamiltonian of the Rindler wedge, evaluating it on the $v>0, u=0$ null Cauchy slice, and doing an inversion $v \to v^{-1}$ to bring $v$ to a finite range. This is a global conformal transformation and leaves the vacuum modular Hamiltonian invariant. 

The Rindler wedge vacuum modular Hamiltonian in arbitrary dimensions is
\bne \label{eq:kri} 
    K = 2\pi H_\text{Rind.} = 2\pi \int_{\Sigma}\dd^{d-1} x \sqrt{h}\,n^\mu k^\mu T_{\mu\nu} 
\ene
where $k = x \del_t + t \del_x$ is the Killing vector field generating boosts in Minkowski spacetime and time translation in Rindler spacetime, $h$ is the induced metric on $\Sigma$, and $n^\mu$ is the normal to $\Sigma$. $\Sigma$ can be any Cauchy slice of the Rindler wedge. Evaluating~\eqref{eq:kri} on the slice $t = \lambda x$, with $|\lambda| \leq 1$, 
gives
\bne \begin{split}
    K &= 2\pi \int \dd^{d-2}x_{\perp} \int_0^\infty \dd v\,v \left(T_{vv} + 2 \frac{1-\lambda}{1+\lambda} T_{uv} + \left( \frac{1-\lambda}{1+\lambda} \right )^2 T_{uu} \right) \\
    &\stackrel{\lambda \to 1}{\sim} 2\pi \int \dd^{d-2}x_{\perp} \int_0^\infty \dd v\,v T_{vv}~.
\end{split}
\label{eq:vacModHam}\ene	
	
For 2d, the last steps to get our desired result~\eqref{eq:mod} are to do a shift $v \to v+1$, an inversion $v \to v^{-1}$, which takes the interval $[1,\infty]$ to $[0,1]$, then a shift and a rescaling to get to the interval $[a,b]$.

 \paragraph{NB on higher dimensions:}
The $v \to v' = v^{-1}$ step we took is only a conformal transformation in two dimensions; the formula for an inversion in arbitrary dimensions is 
\bne 
    x^\mu \to x'^\mu = \frac{x^\mu}{x^2}~.
\ene
One can perform a $v \to v+1$ shift and then inversion of~\eqref{eq:vacModHam} in $d>2$, but only the $u = 0$ null plane remains null after this inversion, and, even for this plane, unlike for $d=2$, we do not get a compact $v$ range after inversion.  

As mentioned earlier, free theories in $d>2$ display ultralocality when quantized on a lightsheet and the theory decomposes into 2d CFTs along each pencil \cite{Wall:2011hj}, to each of which we can apply the inversion $v \to v' = v^{-1}$. So,~\eqref{eq:mod} (with an additional integration over $\int\dd^{d-2}y_\perp$) in $d>2$ is correct for free theories, but not expected to be true for interacting theories. In Section \ref{sect:highd}, we will return to the modular Hamiltonian of a null strip in interacting theories, which can be calculated using that $\Delta K = \Delta S$~\cite{Koeller:2015qmn}.  

\subsubsection{Back to entanglement entropy derivative bounds} \label{sec:BTEEDB}
$B(v)$ is the null interval $[v_B (v),v]$, with $v_B (v) \leq v$ but otherwise arbitrary.
Then the upper bound on $\del_v S(v) = \del_v S(A(v))$ in~\eqref{eq:magtb} becomes
\bne 
\begin{split} \label{eq:upb}
    \del_v S(v) &\leq \rightderiv \Delta K_B + \rightderiv S_\Omega (B) \\
    &=  2\pi \int_{v_B (v)}^v \dd v' \left( \frac{v'-v_B (v)}{v - v_B (v)}\right)^2\langle T_{vv} (v')\rangle+  \frac{\cUV}{6}\frac{1}{v-v_B (v)} \end{split}\ene

Similarly, $C = [v,v_C(v)]$, with $v_C(v) > v$, and the lower bound on $\del_v S(v)$ becomes
\bne\begin{split} \label{eq:downb}
    \del_v S(v) &\geq \leftderiv \Delta K_C + \leftderiv S_\Omega (C) \\
    &= -2\pi \int_v^{v_C(v)} \dd v' \left( \frac{v'-v_C(v)}{v - v_C(v)}\right)^2\langle T_{vv} (v') \rangle - \frac{\cUV}{6}\frac{1}{v_C(v) - v}
\end{split}\ene

Let us focus on the first inequality, eq.~\eqref{eq:upb}, though the analysis we will do can easily be repeated for~\eqref{eq:downb}. The inequality~\eqref{eq:upb} is true for any $v_B (v) \leq v$. The inequality becomes trivial in the limit $v-v_B (v) \to 0^+$,
and, when $v_B (v) \to -\infty$, becomes the half-averaged null energy condition (HANEC): 
\bne 
    2\pi \int_{-\infty}^v \dd v' \langle T_{vv}(v') \rangle \geq  \del_v S(v)~.
\ene
\eqref{eq:downb} also gives HANEC, in the $v_C(v) \to \infty$ limit.

We want the tightest possible bound on $\del_v S(v)$, which is found by minimising the right-hand side of~\eqref{eq:upb} over $v_B (v)$. There is a competition between the two terms: the second term is minimised as $v_B (v) \to -\infty$ (it wants $B$ to be large), but it is counteracted by the first term when the null energy is positive, which wants a small $B$ interval.

We look for the optimal $v_B (v)$ by looking for stationary points. Extremising the RHS of~\eqref{eq:upb} with respect to $v_B (v)$ gives the integral equation
\bne 
    \frac{1}{(v-v_B (v))^2}\int_{v_B (v)}^v \dd v'\left(\frac{24\pi}{\cUV}(v-v')(v'-v_B (v))\langle T_{vv}(v')\rangle-1\right)=0 ~.\label{eq:zesta}	
\ene

Formally, the optimal $v_B (v)$ is the solution to~\eqref{eq:zesta} that minimises the RHS of~\eqref{eq:upb}. But we cannot solve~\eqref{eq:zesta} for arbitrary $\langle T_{vv}\rangle$. Even if we could, there is also the problem of local versus global minima; for example, a simpler version of the problem,
\bne 
    \min_{v_B (v)} \int_{v_B (v)}^v \dd v' \langle T_{vv}(v')\rangle~,
\ene
can have many local extrema -- whenever $\langle T_{vv} (v_B (v))\rangle = 0$ -- so optimising $v_B (v)$ requires global information about $\langle T_{vv} \rangle$. We cannot find the optimal $v_B (v)$ for arbitrary $\langle T_{vv} \rangle$.

However, for some cases,
 one can find an approximate minimising $v_B (v)$. 
If there is an interval in
 $\langle T_{vv} (v) \rangle$ is positive and slowly varying, then the optimal $v_B (v)$, when both $v$ and $v_B$ are within this interval, is
\bne 
    v-v^{(\text{min.})}_B (v) \approx \sqrt{\frac{\cUV}{4\pi \langle T_{vv} (v) \rangle}} . \label{eq:const_T} 
\ene
However, intervals where $\langle T_{vv}(v) \rangle \geq 0$ are not of primary interest for quantum energy inequalities.
But, if $\langle T_{vv}(v) \rangle$ is negative for $ v\in [v_0 , v_1]$, then~\eqref{eq:zesta} has no solution for $v, v_B \in [v_0,v_1]$. The optimal $v_B (v)$ is outside of that interval: $v_B^{\text{(min.)}}(v) \leq v_0$. The optimal $v_B (v)$ is always in a region where $\langle T_{vv} \rangle \geq 0$. Such a region is guaranteed to exist if $\langle T_{vv} \rangle \neq 0$ anywhere, by ANEC.

\subsubsection{\boundName[s]
}\label{sec:SNEC_bounds}
The last step is to bring everything together. From looking at the RHS of~\eqref{eq:integratedQNEC}, given that we restricted to $g(v)$ satisfying $- \text{sgn}(v)g'(v) \leq 0$, we see that we need the upper bound~\eqref{eq:upb} involving $v_B (v)$ on $\del_v S(v)$ for $v<0$, and the lower bound~\eqref{eq:downb} involving $v_C (v)$ for $v\geq 0$. So, for $v<0$, we only need $v_B (v)$, and for $v\geq 0$, we only need $v_C (v)$. Thus, 
we can combine $v_B$ and $v_C$ into a new function 
\bne 
    \zeta(v) = \begin{cases} v_B (v) , &v < 0 \\
    0, & v=0, \\
    v_C (v) , &v > 0~. 
    \label{eq:zetde}
\end{cases} \ene
The key property that $\zeta$ satisfies is
\bne 
    \text{sgn}(v) (\zeta(v) -v) \geq 0 \label{eq:zetac}~.
\ene
Some examples of allowed $\zeta$ include: (1) $v+ \arctan{v}$, (2) $v+ v e^{-v^2}$, (3) $v+\tanh(v)$, and (4) $v+ \frac{v \sin^2(v)}{1+v^2}$. These examples have the additional nice properties that $\zeta$ is odd, invertible, and $|\zeta(v) -v|$ is bounded. 

Next, we plug the bounds~\eqref{eq:upb} and~\eqref{eq:downb} into the RHS of~\eqref{eq:integratedQNEC} and get 
\bne \begin{split} 
    \int_{-\infty}^{\infty} \dd v\,g(v) \langle T_{vv}(v) \rangle &\geq \int_{-\infty}^0 \dd v g'(v) \left (-\int^v_{\zeta(v)} \dd v' \left( \frac{v'-\zeta(v)}{v - \zeta(v)}\right)^2 T_{vv} (v')  - \frac{\cUV}{12\pi}\frac{1}{v- \zeta(v)} \right) \\
    & - \int_0^\infty \dd v g'(v) \left( -\int_v^{\zeta(v)} \dd v' \left( \frac{v'-\zeta(v)}{v - \zeta(v)}\right)^2 T_{vv} (v')  - \frac{\cUV}{12\pi}\frac{1}{\zeta(v) - v} \right) .
\label{eq:longq}
\end{split} \ene
Assuming the integrals commute, and that $\zeta$ is invertible\footnote{For $\zeta (v)$ to be invertible, for invertible $v_B (v)$ and $v_C (v)$, requires $v_B (0) = v_C (0) = 0$. If this condition is not met, the integration orders can still be switched; however, extra contributions are generated. We examine this example in Appendix \ref{sec:non_invertible_zeta}.}%
, we switch the order of integration and simplify to get

\bne \label{eq:not_even_my_final_form}              \boxed{\int_{-\infty}^{\infty} \dd v\, (g(v) +      h(v)) \langle T_{vv}(v) \rangle \geq  -\frac{\cUV}{12\pi} \int_{-\infty}^\infty \dd v  \frac{g'(v)}{v-\zeta(v)} } \ene
where
\bne 
    h(v) := -\int_{\zeta^{-1}(v)}^{v} \dd v'\, g'(v') \left( \frac{v-\zeta(v')}{v'-\zeta(v')}\right)^2~.
\label{eq:hdefn}
\ene
The function $h(v)$ is non-negative, $h(v) \geq 0$,
 and is even if $\zeta (v)$ is odd. $h$ is not necessarily smooth if $\zeta$ is not smooth. Additionally, noting that 
\bne 
    0 \leq \left( \frac{v-\zeta(v')}{v'-\zeta(v')}\right)^2 \leq 1 , \qquad \text{for }v' \in [\zeta^{-1}(v),v]~, 
\ene
we find, from~\eqref{eq:hdefn}, that 
\bne \label{eq:hleq}
    0\leq h(v)\leq g(\zeta^{-1}(v)) - g(v)~,
\ene 
which tells us that $h$ is bounded ($h(v) \leq g(0)$), that $\lim_{|v| \to \infty} h(v) = 0$, and that $h$ has compact support if $g$ has compact support. 

We can get ANEC from the \boundName~\eqref{eq:not_even_my_final_form} by taking constant $g$ (for which $h = 0$ for any $\zeta$). 
$h(v)$ also vanishes if we choose $\zeta(v) =v$, but then the bound~\eqref{eq:not_even_my_final_form} becomes trivial because the RHS diverges. For generic $g(v)$, with $g'(v) \neq 0$ except at $v=0$, $\zeta(v) = v$ is the only choice for which $h(v) = 0$; for it to vanish would require $g'(v')$ to have both signs in the integration range, but it does not because we restricted to $g(v)$ with $-\text{sgn}(v) g'(v) \geq 0$, and the endpoints of the integration range, $v$ and $\zeta^{-1} (v)$, have the same sign, because of~\eqref{eq:zetac}. 

We can do a perturbative analysis. Let $\zeta (v) = v + \epsilon \eta(v)$, with $\eta$ any smooth, bounded function obeying $\epsilon \sgn(v) \eta (v) \geq 0$ (so that~\eqref{eq:zetac} is satisfied).
This gives $\zeta^{-1}(v) = v- \epsilon \eta(v) + O(\epsilon^2)$ and we find $h(v) = - \frac{\epsilon}{3} \eta(v) g'(v) + O(\epsilon^2)$.  
Then the \boundName~\eqref{eq:not_even_my_final_form} becomes
\bne 
    \int_{-\infty}^{\infty} \dd v \,g(v) \left(1+\frac{\epsilon}{3} \eta' (v) + O(\epsilon^2)\right )  \langle T_{vv}(v)\rangle \geq  -\frac{\cUV}{12\pi\epsilon} \int_{-\infty}^\infty \dd v  \frac{(-g'(v))}{\eta(v)}~. \label{eq:perte}
\ene

Next, to explore an example, let us choose $\zeta (v) = av$, with $a\geq 1$ to satisfy~\eqref{eq:zetac}. If we take $g$ to be a normalised Gaussian of width $\sigma$, \eqref{eq:normGauss}, then we get
\bne \label{eq:haexact}
    h(v) = \frac{\left [ -a^2 \,\sigma^2 e^{-\frac{v^2}{2\sigma^2}}
   -\sqrt{2 \pi } a\, \sigma\, v \,\text{erf}\left(\frac{v}{\sqrt{2}\sigma}\right)  +\frac{1}{2} v^2  \text{Ei}\left(-\frac{v^2}{2\sigma^2}\right)  \right]^v_{v/a} }{\sqrt{2 \pi
   } (a-1)^2 \sigma^3
   } 
\ene
where Ei is the exponential integral function.

There are two interesting limits. If we take the limit $a \to\infty$, then we get 
\bne 
    g(v) + h(v) = \frac{1}{\sqrt{2\pi}\sigma} + O(a^{-1})~,
\ene
and~\eqref{eq:not_even_my_final_form} becomes ANEC. 
In the $a \to 1^+$ limit, we can connect this example to~\eqref{eq:perte}, with $\epsilon = a-1$ and $\eta(v) = v$,
for which~\eqref{eq:perte} becomes
\bne \begin{split} 
    \int_{-\infty}^\infty \dd v \,g(v) \left( 1+ \frac{a-1}{3} + O((a-1)^2) \right) \langle T_{vv}(v)\rangle 
    &\geq -\frac{\cUV}{12\pi(a-1)}\int_{-\infty}^\infty dv \,\frac{(-g'(v))}{v} \\
    &= -\frac{c}{12\pi(a-1)\sigma^2}~.
\end{split} \ene
where in the last line we evaluated this for the Gaussian \eqref{eq:normGauss}.

If we compare this to the FH bound~\eqref{eq:fhgau}, we see that, for $a \ll 1$ but finite, this bound is of a similar form, but it is a weaker bound. Since the FH bound is tight, at least for 2d CFTs, we expect that the bounds that we have derived are, in general, weaker than the FH bound. However, it is unclear if this is strictly so or whether, by a clever choice of $\zeta$, we can match the FH bound.

We can put this bound in an alternative form by defining a new smearing function $m(v) := g(v) + h(v)$, and rewrite~\eqref{eq:not_even_my_final_form} as
\bne
    \boxed{\int_{-\infty}^{\infty} \dd v \; m(v) \langle T_{vv}(v)\rangle \geq  -\frac{\cUV}{12\pi} \int_{-\infty}^\infty \dd v \frac{g'(v)}{v-\zeta(v)}}  \label{eq:infSN}
\ene
where
\bne \label{eq:mdef}
    m(v)= g(v)+h(v) = \int_{\zeta^{-1}(v)}^{v} \dd v'\, g(v') \frac{\dd}{\dd v'} \left( \frac{v-\zeta(v')}{v'-\zeta(v')}\right)^2 ~.
\ene
Based on the properties of $h(v)$ in \eqref{eq:hleq}, we find the following relation between $m(v)$ and the original smearing function $g(v)$:
\beq\label{eq:mleq}
    g(v)\leq m(v)\leq g(\zeta^{-1}(v))\qquad\Leftrightarrow\qquad m(\zeta(v))\leq g(v)\leq m(v)~.
\eeq
This implies that $m(v)$ is also non-negative and has compact support when $g(v)$ has compact support. Then, integrating \eqref{eq:not_even_my_final_form} by parts and replacing $g(v)$ with either $m(\zeta(v))$ or $m(v)$ 
(depending on if $\zeta'(v)-1$ is locally postive or negative, respectively)
implies a simpler, yet weaker, form of the bound as
\beq\label{eq:simplified2dbound}
    \boxed{\int dv\,m(v)\,\langle T_{vv}(v)\rangle \geq -\frac{\cUV}{12\pi} \int\dd v\,\frac{m'(\zeta(v))\zeta'(v)\Theta(\zeta'(v)-1) + m'(v)\Theta(1-\zeta'(v))}{v-\zeta(v)}}
\eeq
where $\Theta$ is the Heaviside function. To be clear, we have derived \eqref{eq:simplified2dbound} from a particular convolution of our original smearing function, and it is not clear to us that this implies that $m(v)$ can be treated as a generic smearing function. Regardless, it is tempting to conjecture that \eqref{eq:simplified2dbound} holds for generic smearing functions, $m(v)$ and invertible $\zeta(v)$.

Before concluding this section, we take a moment to re-emphasize that our \boundName[s] are true for any $\zeta (v)$ satisfying its defining property~\eqref{eq:zetac}, and $\zeta(0) = 0$. So,~\eqref{eq:infSN} is an \textit{infinite} family, a function's worth, of 2d \boundName[s]. The choice of $\zeta(v)$ affects both the smearing function $m(v)$ and the lower bound. 

In App.~\ref{sec:testing_bounds}, for a subset of the Ba\~nados states, we verify that some of the important bounds in this subsection are satisfied. In App.~\ref{sec:non_invertible_zeta}, we will relax a simplifying assumption that we made, that $\zeta$ is invertible, 
and so generalise the 2d \boundName[s] of this subsection. The simplest non-invertible $\zeta(v)$ is constant $|\zeta(v) -v|$, i.e., with $\delta >0$,
\bne 
    \zeta(v) = \begin{cases} v-\delta , &v < 0 \\
    0, & v=0, \\
    v+\delta , &v > 0~.
\end{cases} \ene

\section{Higher-dimensional quantum null energy inequalities}\label{sect:highd}
In this section, we will generalize our two-dimensional approach to derive new \boundName[s] in $d>2$ dimensions. 

\subsection{QNEC in higher dimensions}
Consider a $d$-dimensional Minkowski spacetime
\bne 
    \dd s^2 = -\dd u\,\dd v + \dd y_{\perp}^2~,
\ene 
with $\dd y_{\perp}^2 = \sum_{i=1}^{d-2} \dd y_i^2$, and a one-parameter family of regions whose entanglement cuts intersect a constant $u$ null plane at $v = V_\lambda (y_\perp) = V_0 (y_\perp) + \lambda \dot V (y_\perp)$. This gives us a one-parameter family of entropies, $S(\lambda)$. 
Then the QNEC in its general non-local form is
~\cite{Bousso:2015mna, Casini:2022rlv}:
\bne \label{eq:nlq}
    2\pi \int \dd^{d-2} y_\perp \left< T_{vv} (u,v = V_0(y), y_\perp ) \right> \dot{V}(y)^2 \geq \frac{\dd^{2} S} {\dd \lambda^{2}}~. 
\ene
$\frac{d^2 S}{d\lambda^2}$ is an ordinary derivative. It can be written in terms of functional derivatives as follows:
\bne 
    \frac{\dd^{2} S} {\dd \lambda^{2}} = \int \dd^{d-2} y_\perp \dd^{d-2} y^{\prime}_\perp \ \frac{\delta^{2} S} {\delta V ( y_\perp ) \delta V ( y^{\prime}_\perp )}\bigg |_{V_0} \dot{V} ( y_\perp ) \dot{V} ( y^{\prime}_\perp ) ~.
\ene 
The second-order variation of $S$ can be decomposed into a diagonal piece and an off-diagonal piece:%
\footnote{For the reader uncomfortable with functional differentiation, we review the basics and explain this formula in Appendix \ref{sec:func_diff}.}
\bne \label{eq:2dv}
    \frac{\delta^{2} S} {\delta V ( y_\perp ) \delta V ( y^{\prime}_\perp )}=S_{v v}^{\prime\prime} ( y_\perp ) \delta^{( d-2 )} ( y_\perp-y^{\prime}_\perp )+\left( \frac{\delta^{2} S} {\delta V ( y_\perp ) \delta V ( y^{\prime}_\perp )} \right)_{\mathrm{o d}}~. 
\ene
The off-diagonal piece in~\eqref{eq:2dv} is the so-called entanglement density and is non-positive by strong subadditivity~\cite{Faulkner:2015csl}. We can discard the off-diagonal part of~\eqref{eq:2dv} by taking a delta function localised shape deformation
$ \dot{V}(y_\perp)^2 = \delta^{(d-2)}(y_\perp)$.
For such a localised shape deformation, QNEC~\eqref{eq:nlq} reduces to its local form:
\bne \label{eq:localQNEC}
    2\pi \langle T_{vv}  \rangle \geq S''_{vv}. 
\ene
This local form of the QNEC is saturated in interacting CFTs with a twist gap~\cite{Leichenauer:2018obf,Balakrishnan:2019gxl}. 

Next, let us return to QNEC in its general form, eq.~\eqref{eq:nlq}, and consider a family of null-translated flat entanglement cuts: $V_0(y_\perp) = 0$ and $\dot{V}(y_\perp) = 1$. Each entanglement cut in the family is uniform in the transverse directions and parameterised by $\lambda = v$.
Then we can integrate~\eqref{eq:nlq} against $g(v)$, and integrate by parts, to get
\bne 
    2\pi \int \dd v\,g(v) \int \dd^{d-2} y_{\perp} \left< T_{vv} (u,v, y_{\perp}) \right>  \geq - \int \dd v\,g'(v) \frac{\dd S} {\dd v}. \label{eq:intqe}
\ene
This is the higher dimensional version of~\eqref{eq:integratedQNEC}. 

At this point, we could continue to follow the steps of Section~\ref{sect:Rolphbound} to attempt to derive a higher-dimensional \boundName. But the attempt is doomed: no such bound exists. We have already hinted at this in our discussion at the end of that section, namely that within the context of free field theories, it is possible to construct a series of squeezed states with no lower bound on their null-energy density integrated along a single null geodesic. Also, as pointed out in \cite{Fliss:2021gdz}, further integrating this energy density along the transverse directions does not change this conclusion. In addition to the concrete counterexamples in free theory, there exists a general argument against a state-independent lower bound on null energy density integrated over a finite segment of a null geodesic which goes as follows.


Firstly, the operator $\mathcal T=\int \dd v\,g(v) T_{vv}$ with $g$ of compact support is not positive semidefinite, because there exist regions that are spacelike separated from a finite segment of the null line: for an operator $\mathcal{T}$ to be positive semidefinite, it has to annihilate the vacuum. If $\mathcal{T}$ is a local operator smeared over a spacetime region $U$ and it annihilates the vacuum, then it annihilates all states $O_V \ket{0}$, with $O_V$ the set of operators in $V$, and $V$ a region spacelike separated from $U$. But, by the Reeh-Schlieder theorem \cite{Reeh:1961ujh}, the set of states $O_V \ket{0}$ is dense in the Hilbert space, so $\mathcal{T} = 0$. Secondly, if $\mathcal T$ is not positive semidefinite, we can then make it arbitrarily negative by a combination of rescaling and boosts that rescale $T_{vv}$ but leave $g(v)$ unchanged \cite{WittenTalk}. To be more specific, since $T_{vv}$ is a primary under dilatations, there exists a unitary transformation $U_{D(\lambda)}$ such that
\beq
    T_{vv}(\lambda u,\lambda v,\lambda \vec y_\perp)=\lambda^{-d}U_{D(\lambda)}T_{vv}(u,v,y_\perp)U^\dagger_{D(\lambda)}~.
\eeq
Similarly, under boosts there exists a unitary operator $U_{K(\lambda)}$ such that
\beq
    T_{vv}(\lambda u,\lambda^{-1} v,\vec y_{\perp})=\lambda^{2}U_{K(\lambda)}T_{vv}(u,v,y_\perp)U^\dagger_{K(\lambda)}~.
\eeq
Since $\int\dd v\,g(v)\,T_{vv}$ is not a positive operator, we know there exists a state $\ket{\psi}$ such that
\beq
    \int\dd v\,g(v)\,\langle T_{vv}(0,v,\vec 0)\rangle_{\psi}= - \mc F~,
\eeq
for some positive quantity $\mc F$. Then it's clear that under a combination of dilatation and boost preserving the null line, there then exists a state $\ket{\tilde\psi}=U_{K(\lambda)}U_{D(\lambda)}\ket{\psi}$ such that
\beq
    \int\dd v\,g(v)\langle T_{vv}(0,v,\vec 0)\rangle_{\tilde\psi}=-\lambda^{d-2}\mc F~.
\eeq
Then $\int\dd v\,g(v)\,\langle T_{vv}\rangle$ can be made arbitrarily negative and so cannot admit a state-independent lower bound.

From \eqref{eq:intqe} we can alternatively track the triviality of any putative \boundName from a divergence appearing in the derivative of the entanglement entropy associated with a null interval. We can reinterpret the conclusions of \cite{Fliss:2021gdz} for free theories from this perspective as well. For instance, consider the free bosonic theory quantized on a lightsheet $\mc L$. In the procedure we developed in Section \ref{sect:Rolphbound}, our lower bounds were set by derivatives of the entanglement entropy reduced on a null interval, and so we consider here a null strip 
with $v\in [v_1,v_2]$. As emphasized in \cite{Wall:2011hj}, the vacuum factorizes into the vacuum along a coarse-grained null ray, or ``pencil'' $\mathfrak p$, 
as
\beq
    |\Omega\rangle_{\mathcal L}=\bigotimes_{\mathfrak p}|\Omega\rangle_{\mathfrak p}~,
\eeq
where each $|\Omega\rangle_{\mathfrak p}$ is the vacuum of a 2d chiral scalar CFT. 
Then, it follows that the strip's entanglement entropy is additive to a sum of CFT entropies along each pencil:
\beq
    S_{\Omega}[v_2,v_1]=\sum_{\mathfrak p}S_{\Omega_\mathfrak p}[v_2,v_1]=\frac{1}{6}\sum_{\mathfrak p}\log(v_2-v_1)~.
\eeq
In the limit that the pencils are very narrow, this sum is replaced by the integral 
\beq
    S_\Omega[v_2,v_1]=\frac{1}{a^{d-2}}\int \dd^{d-2}y_\perp\,\left(\frac{1}{6}\log(v_2-v_1)\right)~,
\eeq
where $a$ is the pencil width. Considering, for instance, $\rightderiv S_\Omega$, we see it is both UV and IR divergent:
\beq
    \rightderiv S_\Omega\equiv\frac{\dd}{\dd v_2}S_\Omega=\frac{1}{6a^{d-2}}\,\frac{V_\perp}{v_2-v_1}~,
\eeq
with $V_\perp=\int\dd^{d-2}y_\perp$.

\subsection{A family of interacting \boundName[s]}\label{sec:InteractionDSNEC}

We will return to and leverage the general approach of Section \ref{sect:Rolphbound}. Namely, we will bound the stress tensor in a state, smeared over a function $g(v)$ with compact support\footnote{This assumption of compact support is for the ease of illustration. There should be no difficulty in extending our arguments to Schwartz functions as well.} in a causal 
diamond extended in the $\vec{y}_\perp \in \mathbb{R}^{d-2}$ transverse space, in terms of null derivatives of an entanglement entropy through the integrated QNEC, \eqref{eq:intqe}:
\begin{align}\label{eq:DSNECsetup1}
    \int \dd^{d-2}y_\perp\int\,\dd v\,g(v)\langle T_{vv}(u_0,v,y_\perp)\rangle&\geq\frac{1}{2\pi}\int \dd u \,\dd v\,g(v)\pa_v^2S(A)\nonumber\\
    &=-\frac{1}{2\pi}\int \dd v\,\pa_vg(v)\pa_vS(A)~.
\end{align}
Here $S(A)$ is the entanglement entropy of the state $\psi$ reduced on a region $A$, which is any achronal codimension-1 half-space whose boundary $\del A$ is a flat cut at $(u_0,v)$ for all $\vec y_\perp\in\mathbb R^{d-2}$:
\beq
    \pa A(v) =\Big\{(u_0,v,y_\perp)\Big|y_\perp\in\mathbb R^{d-2}\Big\}~.
\eeq
The region $A$ depends on the value of $v$, and there is a different region for each point in the integral appearing in \eqref{eq:DSNECsetup1}. The difference between this set-up and that of Section \ref{sect:Rolphbound} is that in bounding the first derivatives of $S(A)$, to regulate the UV divergences, we will take our strip regions to not be exactly null, but nearly null. 

To be explicit, let us illustrate the upper bound on $\pa_vS(A)$. Let $B$ be an achronal, codimension-1 strip of constant width (i.e. $B$ is an interval times $\mathbb R^{d-2}_{\vec y_\perp}$) sharing one of its two boundary components with $A$:
\beq
    \pa B=\pa A\cup\Big\{(u_B,v_B,\vec y_\perp)\Big|\vec y_\perp\in\mathbb R^{d-2}\Big\}~,
\eeq
and whose domain of dependence is contained in that of $A$: $\mc D(B)\subset\mc D(A)$. Such a region is illustrated on the left of Fig. \ref{fig:skinnyDSNECsetup}. Much like the region $A$, $B$ must depend explicitly on the point $(u_0,v)$ in the $u$-$v$ plane, at the very least, by virtue of sharing a common boundary with $A$; we can also use this freedom to choose the second boundary at $\{(u_B,v_B,\vec y_\perp)\}$ to vary as we vary $v$, i.e. $(u_B,v_B)\rightarrow (u_B(v),v_B(v))$. However in order to maintain $\mc D(B)\subset\mc D(A)$ we must have $u_B(v)>u_0$ and $v_B(u,v)<v$ for all $v$ of interest. 
We will see major simplifications in the limit that $B$ is taken to be nearly null:
\beq
\epsilon_B:=u_B-u_0\ll1~.
\eeq

For the lower bound, we will consider a region $C$ to be a nearly null strip 
\bne 
    \del C = \del A \cup \Big\{ (u_C,v_C,\vec{y}_\perp) \Big|\vec{y}_\perp \in \mathbb R^{d-2}\Big\}~.
\ene
Again, $C$ depends on $v$ by sharing a boundary with $A$, and we can additionally allow its right boundary to vary with $v$ as well. However we will maintain $u_C(v)<u_0$ and $v_C(v)>v$ such that domain of dependence of $C$ is contained in the domain of the complement of $A$: $\mc D(C)\subset \mc D(A^c)$, as depicted in the right of Figure \ref{fig:skinnyDSNECsetup}. 
Again, we will want to take a limit that $C$ becomes null, which will require
\beq
\epsilon_C:=u_0-u_C\ll 1
\eeq
for all $v$.

\begin{figure}[h!]
\centering
\begin{tikzpicture}
    \begin{scope}[rotate=45]


    \end{scope}

    \draw[thick,dashed] (-1.5,-1.5) to (1.5,1.5);
    \draw[thick,dashed] (1.5,-1.5) to (-1.5,1.5);
    \draw[thick,blue] (-2.5,0) to (0,0);
    \draw[thick,orange] (-2,-1.2) to (0,0);

    \node[blue] at (-1.75,.25) {$A$};
    \node[orange] at (-1.5, -.65) {$B$};

    \draw[fill] (0,0) circle (1pt) node [right] {$~(u_0,v)$};
    \draw[fill] (-2,-1.2) circle (1pt) node [left] {$(u_B,v_B)$};
\end{tikzpicture}
\qquad\qquad
\begin{tikzpicture}
    \draw[thick,dashed] (-1.5,-1.5) to (1.5,1.5);
    \draw[thick,dashed] (1.5,-1.5) to (-1.5,1.5);
    \draw[thick,blue] (-2.5,0) to (0,0);
    \draw[thick,orange] (0,0) to (1.4,.8);
    \node[blue] at (-1.75,.25) {$A$};
    \node[orange] at (.55, .75) {$C$};
    \draw[fill] (0,0) circle (1pt) node [right] {$~(u_0,v)$};
    \draw[fill] (1.4,.8) circle (1pt) node [right] {$(u_C,v_C)$};
\end{tikzpicture}
\caption{The $v$-dependent family of codim-$1$ regions $A$, $B$, and $C$ in the setup for the higher dimensional derivation, all of which extend uniformly in the transverse $\mathbb R^{d-2}$ directions. $A$ is a half-plane and $B$ and $C$ are strips.
}
\label{fig:skinnyDSNECsetup}
\end{figure}

Much of what we established in Section \ref{sect:Rolphbound} still holds. Namely, since $\mc D(B)\subset\mc D(A)$ and $\mc D(C)\subset \mc D(A^c)$ we have
\beq
    \leftderiv S(C)\leq\frac{\dd}{\dd v}S(A)\leq\rightderiv S(B)~.
    \label{eq:twodb}
\eeq
As in $d=2$ (see~\eqref{eq:righd} and~\eqref{eq:leftd}), $\leftderiv$ and $\rightderiv$ denote that $\del_v$ acts only on the boundary shared with $A$.
In this two-sided bound, all that has changed with respect to $d =2$ is that $A$, $B$ and $C$ have an extension into the transverse space $y_\perp \in \mathbb{R}^{d-2}$. But, in $d>2$,
the derivatives of the vacuum entropies are not UV-finite when we take the null limit of the strip regions, so we will regulate by taking $B$ and $C$ to be nearly null. However, when $B$ and $C$ are not strictly null, we lose some of the universality of the results of Section \ref{sect:Rolphbound}. In particular, we do not have a closed-form expression for entanglement entropies of $S(B)$ and $S(C)$, nor local expressions for the vacuum-subtracted modular Hamiltonian.
Nonetheless, we can work perturbatively in the nearly-null limit. In particular, we can still express
\beq
    S(B)=\Delta S(B)+S_\Omega(B)~,
\eeq
with $\Delta S(B)$ finite in null limit $\epsilon_B\rightarrow 0$; all divergences instead arise from the vacuum entropy. Similar statements apply to $S(C)$. 

To see why this is so, we evaluate the entanglement entropy through the replica trick. Assuming that our state in question can be prepared by path-integral preparation with operator insertions\footnote{By the state operator correspondence, such states are dense in a CFT.} we can express the entanglement entropy as
\beq
    S(B)=\lim_{n\rightarrow1}\frac{1}{1-n}\log\frac{Z_n}{Z_1^n}~,
\eeq
where $Z_n$ the CFT path-integral on the $n-$fold branched cover, $M_n$, with the insertions preparing $\psi$ replicated on each branch. We can map this back to a partition function of a symmetric product CFT, which we denote as $(\text{CFT})^n/\mathbb Z_n$, on a trivial topology. The modding by $\mathbb Z_n$ means we only allow replica symmetric operators in the spectrum.\footnote{As emphasized in \cite{Balakrishnan:2019gxl}, this projection is necessary for the defect CFT living on the entanglement cut to be a local CFT.} The replica trick is encoded by the insertion of twist-1 
codim-2 defect operators $D_n$ at the boundaries of the strip:
\beq
    S(B)=\lim_{n\rightarrow 1}\frac{1}{1-n}\log\langle D_n(u_B,v_B)D_n(u_0,v)\rangle~.
\eeq
The twist defect operators admit an operator product expansion (OPE) of the following form\footnote{The coefficient of $(1-n)$ in the exponential is assumed based on the assumption of analyticity of the R\'enyi entropy in $n$~\cite{Headrick:2010zt}.} \cite{Bousso:2014uxa}
\beq\label{eq:DOPEschem}
    D_n(u,v)D_n(0)\sim\exp\left(-(n-1)\int\dd^{d-2}y_\perp\left[\sum_{\mc O}|x|^{-(d-2)+\Delta_{\mc O}}\mc O(0,y_\perp)\right]\right)
\eeq
with $|x|^2=|uv|$. 
The sum over $\mc O$ in \eqref{eq:DOPEschem} is a sum over replica symmetric primary operators of $(\text{CFT})^n/\mathbb Z_n$ which includes `single-copy' operators of the `seed' CFT, and can be expressed in terms of primary operators of the `seed' CFT as
\beq\label{eq:repsymmOschem}
    \sum_{i=0}^{n-1}\mc O^{(i)}~,
\eeq
where $\mc O^{(i)}$ belongs to the $i^\text{th}$ CFT, as well as double-copy $\sum_{i,j}\mc O^{(i,j)}$ operators, and multi-copy operators. 
We will soon see that our nearly null limit will entail that we only concern ourselves with the single-copy operators of \eqref{eq:repsymmOschem}, however, we will briefly discuss the contribution of double-copy operators in the context of free fields below.

The lightcone limit $u\rightarrow0$ of \eqref{eq:DOPEschem} is very similar to a coincident limit of this OPE and takes a simplified form \cite{Casini:2017roe}
\beq\label{eq:nullDOPEschem}
    \lim_{u\rightarrow 0}D_n(u,v)D_n(0)=\exp\left(-(n-1)\sum_{\mc O}\int \dd^{d-2} y_\perp \int_0^{v}\dd v'\frac{v^{\prime~s-1}}{|uv^\prime|^{\frac{1}{2}(d-2-\tau_{\mc O})}}\mc O_{\Delta,s}(u,v',y_\perp)\right)
\eeq
where $\tau_{\mc O}=\Delta_{\mc O}-s_{\mc O}$ is the twist of the operator $\mc O$ with respect to boosts about the entangling surface. In particular, the most singular terms of \eqref{eq:nullDOPEschem} come from operators with minimum twist. In this paper, we will make the assumption of a {\it twist gap}, namely aside for the identity operator with $\Delta_1=s_1=\tau_1=0$, the operator with lowest twist is the stress tensor itself with $\tau_T=d-2$.
Namely, the leading singular and non-singular contributions to \eqref{eq:nullDOPEschem} are
\begin{align}\label{eq:DDnulllimit}
    D_n(u,v)D_n(0)=\exp \Bigg[ (1-n) \int_{\vec y_\perp} &\Bigg(\frac{2\beta_d\,\cT}{d-2}\frac{\hat{\mathbbm 1}}{|uv|^{\frac{d-2}{2}}}\nonumber\\
    &\qquad+\int_0^{v}\dd v'\,v'\left(\sum_{i=0}^{n-1}T_{vv}^{(i)}(u,v',y_\perp)+\text{desc.}\right)+O(u)\Bigg)\Bigg]~,
\end{align}
where we have expressed the replica symmetric operator explicitly as a sum over seed stress tensors and indicated the sum over its descendants schematically. We will make the additional assumption that we can expand this OPE about $n = 1$ to arrive at 
\begin{align}
D_n(u,v)D_n(0)= 1&-(n-1) \int_{\vec y_\perp}\left(\frac{2\beta_d\,\cT}{d-2}\frac{\hat{\mathbbm{1}}}{|uv|^{\frac{d-2}{2}}}+\int_0^{v}\dd v'\,v'~\left(T_{vv}(u,v',y_\perp)+\text{desc.}\right)\right)\nonumber\\
&\qquad\qquad\qquad\qquad\qquad\qquad\qquad\qquad+O(u)+O((n-1)^2)~.
\end{align}

Because the VEV of the single-copy stress-tensors vanishes, the identity term of \eqref{eq:DDnulllimit} is the leading contribution to the vacuum entanglement entropy $S_\Omega(B)$. The stress tensor contribution instead determines the entropy difference $\Delta S$. Note that this second term is finite in the null limit. In the strict null limit, $u\rightarrow 0$, this is also equal to the vacuum subtracted modular Hamiltonian, $\Delta S=\Delta K$, due to the vanishing of the relative entropy for interacting theories on a null sheet \cite{Bousso:2014uxa}. From this finiteness, we will assume that it admits an expansion in the separation in the $u$-direction 
\beq
    \Delta S(B)=2\pi \int d^{d-2} y_\perp \int_{v_B}^v\dd v'(v-v_B)G\left(\frac{v'-v_B}{v-v_B}\right)\langle T_{vv}(u_0,v',y_\perp)\rangle+O(\epsilon_B)~,
    \label{eq:kexps}
\eeq
Here $G$ encapsulates the resummation of the descendants in the stress tensor channel of the twist defect OPE. Thus it is theory-specific. For instance, the free theory modular Hamiltonian is given by $G(s)=s(1-s)$ (consistent with what we have seen in Section \ref{sect:Rolphbound}), however is known to differ for holographic CFTs \cite{Bousso:2014uxa}. Regardless, $G(s)$ obeys several properties that were derived in \cite{Bousso:2014uxa}:
\beq\label{eq:Gprops}
    G(0)=G(1)=0~,\qquad G(s)=G(1-s)~,\qquad G'(0)=-G'(1)=1~,\qquad |G'(s)|\leq 1~.
\eeq
It is also conjectured in that paper that $G(s)$ is concave,
\beq
    G''(s)\leq 0~,
\eeq
however this is to date unproven. We will return to this conjectured property later. We thus have
\beq
    \leftderiv\Delta S(B)=2\pi\int d^{d-2} y_\perp \int_{v_B}^v\dd v'\mc G\left(\frac{v'-v_B}{v-v_B}\right)\langle T_{vv}(u_0,v',y_\perp)\rangle+O(\epsilon_B)~,
\eeq
where
\beq
\mc G(s)\equiv G(s)-s\,G'(s)~.
\eeq

We now focus on the divergent and finite contributions to the vacuum entropy. It is natural to assume, as we have above, that the coefficient for the identity operator term in the defect OPE is proportional to the prefactor $\cT$ of the stress-tensor two-point function\footnote{We use the normalization convention of \cite{Osborn:1993cr}, which is most easily stated in a time-like colinear frame as
\beq
\langle T_{00}(t)T_{00}(0)\rangle_\Omega=\frac{(d-1)}{d}\frac{\cT}{t^{2d}}~.
\eeq}. This assumption can be corroborated through the calculation of a nearly null strip in a holographic CFT using the Ryu-Takayanagi formula \cite{Ryu:2006bv}. In this context, the bulk Newton's constant is proportional to $\cT$ \cite{Buchel:2009sk} which fixes the coefficient $\beta_d$ to be 
\beq
    \beta_d=\frac{d-1}{d+1}2^{d-1}\pi^d\frac{\Gamma\left(\frac{d}{2}\right)}{\Gamma(d+1)}\left(\frac{\Gamma\left(\frac{d}{2d-2}\right)}{\Gamma\left(\frac{1}{2d-2}\right)}\right)^{d-1}~,
\eeq
although we acknowledge that this coefficient may differ for non-holographic CFTs; in any case, its precise form will not be important for what follows. In the $\epsilon_B=(u_B-u_0)\rightarrow0$ limit, the defect OPE computation suggests that the vacuum entanglement entropy of a strip in the null limit takes a universal form in interacting CFTs\footnote{It is feasible that a generic interacting QFT might also have significant simplifications in its OPE in the null limit. However, because we are using features of the defect OPE (i.e. convergence of the OPE) that have only been strictly established for CFTs, we will be cautious and specify CFTs in this section.}: 
\beq\label{eq:nearlynullSvac}
   S_\Omega(B)=\frac{2\beta_d\,\cT V_\perp}{d-2}\left(\frac{1}{\epsilon_\text{UV}^{d-2}}-\frac{1}{|\epsilon_B|^{\frac{d-2}{2}}|v-v_B|^{\frac{d-2}{2}}}\right)+O(\epsilon_B)~,
\eeq
where $V_\perp\equiv\int\dd^{d-2}y_\perp$ is the IR-divergent transverse volume. To be careful and clear in this null limit, \eqref{eq:nearlynullSvac} represents the divergent and finite terms in a $\epsilon_B$ expansion while holding $v-v_B$ fixed such that $\epsilon_B|v-v_B|> \epsilon_\text{UV}^2$, where $\epsilon_\text{UV}$ is an inverse UV cutoff. We have additionally allowed for a $(u,v)$ independent divergence with the UV cutoff, $\epsilon_\text{UV}$ stemming from a normalization of the $D_nD_n$ OPE; however, importantly, {\it there are no further subleading divergences in $\epsilon_\text{UV}$.} 
Subleading terms would be functionals of curvature invariants of the entanglement cut,%
\footnote{The general structure of the vacuum EE of any region $\Sigma$ in any local QFT is~\cite{Casini:2022rlv}
\bne S_{vac.} (\Sigma) = \sum C_i (\del \Sigma) \epsilon_\text{UV}^{-\lambda_i} + S_0 (\Sigma) \ene
where $C_i (\del \Sigma)$ have dimension $\lambda_i$ and are local and extensive functionals of the curvature invariants of the boundary $\del \Sigma$:
\bne C_{i} (\del\Sigma) = \int_{\del \Sigma}  (\dots) \ene
}
 which vanish for our flat entanglement cuts.
For holographic CFTs, the vacuum entropy of a strip appeared as an example in the original paper by Ryu and Takayanagi, \cite{Ryu:2006bv}, however, to our knowledge, the universality of this result for general interacting CFTs in the null limit has not yet been emphasized in the literature.

The derivative of $S_\Omega$ is cutoff independent however:
\beq
    \rightderiv S_\Omega(B)=\frac{\beta_d\,\cT\,V_\perp}{\epsilon_B^{\frac{d-2}{2}}(v-v_B)^{\frac{d}{2}}}+O(\epsilon_B)~.
\eeq

\subsubsection*{NB on free theories}

As we mentioned, \eqref{eq:DDnulllimit} supposes that the only operator saturating the twist gap is the single-copy stress tensor, $\sum_iT_{vv}^{(i)}$. However, in free theories, this assumption is not true. For instance, there exists an infinite tower of single-copy higher-spin operators, schematically,
\beq
    \mc J_\text{single}^{(s)}=\sum_{i}\pa_v^{(s/2)}\phi^{(i)}\pa_v^{(s/2)}\phi^{(i)}~,
\eeq
for all even $s$ which can alter the expression for $\Delta S$. In this case, $\Delta S\neq \Delta K$ and we need to instead use the monotonicity of the relative entropy (as in Section \ref{sect:Rolphbound}) to isolate an inequality involving only the stress tensor. Moreover, there are also double-copy higher-spin operators of the form
\beq
    \mc J^{(s)}_\text{double}=\sum_{i,j}\pa_v^{(s/2)}\phi^{(i)}\pa_v^{(s/2)}\phi^{(j)}~.
\eeq
These double-copy operators will have non-zero VEV which will alter the vacuum entropy itself. In the presence of interactions, such operators, with the exception of the single-copy stress tensor itself, will acquire positive anomalous dimensions and will no longer saturate the twist gap.

Returning to our upper and lower entropy bounds, we have
\begin{align}\label{eq:highDSAupperbound}
    \pa_vS(A)\leq \int \dd^{d-2}y_\perp&\left[2\pi\int_{v_B}^v\dd v'\mc G\left(\frac{v'-v_B}{v-v_B}\right)\langle T_{vv}(u_0,v',y_\perp)\rangle+\frac{\beta_d\,\cT}{\epsilon_B(v)^{\frac{d-2}{2}}(v-v_B)^{\frac{d}{2}}}\right]\nonumber\\
    &\qquad+O(\epsilon_B)
\end{align}
A similar result holds for the nearly-null ($\epsilon_C \to 0^+$)  strip region $C$ depicted on the right in Figure \ref{fig:skinnyDSNECsetup}, which establishes a lower bound on the null variation of $S(A)$:
\begin{align}\label{eq:highDSAlowerbound}
    \pa_vS(A)\geq\int\dd^{d-2}y_\perp&\left[-2\pi\int_{v}^{v_C}\dd v'\mc G\left(\frac{v_C-v'}{v_C-v}\right)\langle T_{vv}(u_0,v',y_\perp)\rangle-\frac{\beta_d\,\cT}{\epsilon_C(v)^\frac{d-2}{2}(v_C-v)^{\frac{d}{2}}}\right]\nonumber\\
    &\qquad+O(\epsilon_C)~,
\end{align}
where, using the properties of $G(s)$, \eqref{eq:Gprops}, we've employed $\mc G(1-s)=\mc G(s)+G'(s)~.$

We can now follow similar steps to those in Section \ref{sect:Rolphbound}. Again, we restrict $g(v)$ to be such that $-\text{sgn}(v)\pa_vg\geq 0$, and we define functions
\beq
    \zeta(v)=\begin{cases} v_B (v) , &v < 0 \\
    v_C (v) , &v \geq 0 \\
    \end{cases}~,\qquad \epsilon_u(v)=\begin{cases} \epsilon_B(v) , &v < 0 \\
    \epsilon_C(v) , &v \geq 0 \\
    \end{cases}~.
\eeq
More generally, we can take $\zeta(v)$ and $\epsilon_u(v)$ to be any functions such that
\beq
    \text{sgn}(v)\text{sgn}(\zeta(v)-v)\geq 0~,\qquad -\text{sgn}(v)\text{sgn}(\epsilon_u(v))\geq 0~.
\eeq
Next, we invert the integration domain to arrive at
\begin{align} \label{eq:higherDSNEC}
    \int\dd^{d-2}y_\perp\int \dd v&\mc M(v)\,\langle T_{vv}\rangle\geq\nonumber\\
    &-\frac{\beta_d\,\cT\,V_\perp}{2\pi}\int \dd v\,\left(\frac{\pa_vg (v)}{\epsilon_u(v)^{\frac{d-2}{2}}(v-\zeta(v))^{\frac{d}{2}}}+O(\epsilon_u)\right)~,
\end{align}
with
\begin{align}\label{eq:Mdef}
    \mc M(v):=&g(v)-\int_{\zeta^{-1}(v)}^v\,\dd v'(\pa_{v'}g(v'))\mc G\left(\frac{v-\zeta(v')}{v'-\zeta(v')}\right)\nonumber\\
    =&-\frac{1}{2}\int_{\zeta^{-1}(v)}^v\dd v'\,g(v')\,G''\left(\frac{v-\zeta(v')}{v'-\zeta(v')}\right)\frac{\dd}{\dd v'}\left(\frac{v-\zeta(v')}{v'-\zeta(v')}\right)^2~.
\end{align}
We see that the function $\mc M$ takes a form very similar to the function appearing in our 2d QNEIs \eqref{eq:mdef} with the exception of its defining integral being weighted by the second derivative of the kernel appearing in the modular Hamiltonian associated to a null strip, \eqref{eq:kexps}. Note that for free field theories $G''(s)=-2$, in which case $\mc M$ coincides with $m$. More generally though, because of the theory-dependent nature of this modular kernel, it is hard to establish general properties of $\mc M$. However we do point out that when $G$ is concave, i.e. $G''(s)\leq 0$, as was conjectured in \cite{Bousso:2014uxa}, then we can easily establish that $\mc M$ obeys bounds analogous to \eqref{eq:mleq}, namely
\beq
    g(v)\leq \mc M(v)\leq g(\zeta^{-1}(v))~.
\eeq
These bounds again would imply that $\mc M$ is positive, and has compact support when $g$ has compact support.

Similar manipulations to those at the end of Section \ref{sect:Rolphbound} then allow a simplified bound of the form
\begin{align}\label{eq:simplifiedhighdbound}
    \int\dd^{d-2}y_\perp&\int\dd v\,\mc M(u,v)\langle T_{vv}(u_0,v,\vec y_\perp)\rangle\geq\nonumber\\
    &-\frac{\beta_d \cT V_\perp}{2\pi}\int\dd v\left(\frac{\pa_{v}\mc M(\zeta)\Theta(\pa_v\zeta-1)+\pa_v\mc M(v)\Theta(1-\pa_v\zeta)}{\epsilon_u^{\frac{d-2}{2}}(v-\zeta)^{\frac{d}{2}}}+O(\epsilon_u)\right)
\end{align}
where for brevity of notation we have left the $v$ dependence of $\epsilon_u$ and $\zeta$ implicit.

Again, much like our simplified 2d QEI, \eqref{eq:simplified2dbound}, \eqref{eq:simplifiedhighdbound} is now expressed in terms of a single smearing function, $\mc M$, and two functions, $\zeta$ and $\epsilon_u$ parameterizing choices of nearly null intervals. For a given $G$, which is fixed by the theory in question, it is unclear to us whether any generic smearing function $\mc M$ can be realized by a choice of $g$ appearing in \eqref{eq:Mdef}. If so, then \eqref{eq:simplified2dbound} and \eqref{eq:simplifiedhighdbound} would represent a very tantalizing result: a universal family of \boundName[s] for interacting CFTs which, up to an overall constant, are independent of the particular details of a CFT.

\section{Discussion} \label{sec:disc}

In this paper, we have derived new quantum null energy inequalities, i.e. lower bounds on $\langle T_{vv}\rangle_\psi$, the null-null component of the stress tensor, integrated over a spacetime region in interacting quantum field theories. While our study of QNEIs is partially motivated by their potential applications to semi-classical gravity (e.g. as input in a semi-classical singularity theorem) such lower bounds on null energy are, of course, of intrinsic interest in the study of interacting QFTs. In particular, the foundations of our \boundName[s] are deep connections between energy and entropy: in our derivation, we utilize QNEC, strong subadditivity, and the vacuum modular Hamiltonians of null strips. More specifically, given a positive smearing function, $g$, we bound the stress tensor integrated against a smooth kernel, determined by the properties of the modular Hamiltonian of a null strip, in terms of a state-independent functional of $g$. In the case of $d>2$ dimensions, our bounds are the first such 
state-independent QEIs integrated over a finite null segment in a broad class of interacting QFTs, namely CFTs.

For $d=2$, these QNEIs are 
\beq
    \int\dd v\, m(v)\langle T_{vv}\rangle\geq-\frac{\cUV}{12\pi}\int\dd v\frac{g'(v)}{v-\zeta(v)}~.
\eeq
where a precise form of $m$ can be found in \eqref{eq:mdef}. In contrast to QNEC, these lower bounds are state-independent, because the right-hand side does not depend on the state $\psi$. And, in contrast to ANEC, these bounds are localised, in the sense that we can choose $g$ such that we are integrating $\langle T_{vv} \rangle$ against a bump function on the left-hand side. 

We have an infinite family of QNEIs, one for each choice of $\zeta$, which is both powerful but also increases the complexity of the inequality. The choice of $\zeta$ affects both the definition of $m$, the correction to the smearing function, and the value of the lower bound and arises from a choice of null intervals appearing in intermediate stages of our derivation. However, ultimately $\zeta$ is an invertible function obeying~\eqref{eq:zetac}, but otherwise arbitrary. 
For a given $\langle T_{vv} \rangle$, there does exist a $\zeta$ that makes this intermediate stage inequality as tight as possible, but, for general $\langle T_{vv} \rangle$, there is no analytic solution for this $\zeta$ (see Sec.~\ref{sec:BTEEDB}). 
We explicitly determined the QNEI for a few example choices of $\zeta$. It would be interesting to see whether the general form of our QNEIs can be simplified.

In higher dimensions, the QNEIs are of the form 
\beq
    \begin{split}\int\dd^{d-2}y_\perp&\int\dd v\mc M(v)\langle T_{vv}(u_0,v,y_\perp)\rangle\\
    &\geq-\frac{\beta\,\cT}{2\pi}\int\dd^{d-2}y_\perp\int\dd v\left(\frac{\pa_vg(v)}{\epsilon_u(v)^{\frac{d-2}{2}}(v-\zeta(v))^{\frac{d}{2}}}+O(\epsilon_u)\right)~,
    \end{split}
    \label{eq:higher_dim_QNEI}
\eeq
where $\mc M$ is defined in \eqref{eq:Mdef}. As for $d=2$, this inequality is state-independent,%
\footnote{
To be precise, we have only shown state-independence for powers of $\epsilon_u$ between $\epsilon_u^{-\frac{d-2}{2}}$ and $\epsilon_u^0$.
We expect that the subleading, positive-power orders in $\epsilon_u$ in the bound~\eqref{eq:higher_dim_QNEI} are not state-independent, that they depend on the expectation values of the single and multi-copy operators with twist $\tau>d-2$ (greater than that of the single-copy stress tensor) that appear in the twist-twist OPE~\eqref{eq:DDnulllimit}.}
and the integral over $\langle T_{vv} \rangle$ is localised (in the $(u,v)$ plane). 
Unlike for $d=2$, we integrate $\langle T_{vv} \rangle$ over the transverse directions $y_\perp$. 
Additionally, this bound depends on a second function, $\epsilon_u$, which parametrises how close we take our strip regions to being null in an intermediate step of the derivation (see Sec.~\eqref{sec:InteractionDSNEC}), and we have to work perturbatively in this function, to make use of the universal form of $\Delta K$ for null strips. To our knowledge, these are the first QEIs, integrated over a finite null region, proven for higher-dimensional interacting QFTs.

While we view this work as an important first step in the study of \boundName[s] in interacting theories, there is still much more to explore:
\begin{itemize}
    \item Our \boundName[s] in higher dimensions are localised in the $(u,v)$ plane, but not in the transverse space, so they are not truly localised bounds on the stress tensor. Additionally, this integration strictly trivialises the right-hand side of our bounds (albeit in a physically uninteresting way). It would obviously be interesting and useful to try to leverage the techniques we've established here to find bounds that are also localized in this transverse direction. 
    At a technical level, it was necessary to have a flat entanglement cut in the transverse directions to make use of the results for $\Delta K$ for a null strip given in~\cite{Bousso:2014uxa}. Having the expression for $\Delta K$ for null strips with ``wiggly edges'' would allow us to utilize local variations and the QNEC in its local form, \eqref{eq:localQNEC}. However, this expression for the modular Hamiltonian is not known outside of free theories. The modular Hamiltonians of infinite null half-sheets with wiggly edges {\it do} admit closed form expressions \cite{Faulkner:2016mzt,Casini:2017roe} however we have so far not been able to leverage this expression to derive any new interesting bounds. Regardless, it is worth revisiting and reviewing this approach more carefully in the future.
    We expect that it is possible to localise the QNEI~\eqref{eq:higher_dim_QNEI} in the transverse directions. Removing the integrals over $y_\perp$ in~\eqref{eq:higher_dim_QNEI} gives a reasonable, transversally-localised QNEI conjecture. We leave the proof for future work.
    
    It is also useful to look for alternative approaches for deriving a truly local QNEI. One potential approach is to try to realize the stress tensor in the OPE of two operators. This is essentially a more sophisticated form of the free field theory statement that $T_{vv}$ is the square of two operators. One obvious hang-up to this approach is that operator spectra in CFTs are not universal and so there is potential ambiguity in which OPE we try to realize stress tensor. To address this, we can once again utilize entanglement-based techniques. Namely, {\it defect CFTs} living along twist defects defining replica path-integrals admit a universal set of displacement operators with the stress tensor appearing in their OPE channel \cite{Balakrishnan:2019gxl}. One might speculatively hope that a reflection-symmetric OPE of such displacement operators could provide a universal and local QEI in interacting CFTs. 
    
    \item We have worked in Minkowski spacetime, both for simplicity and because our QNEIs are only proven for theories and background geometries for which our inputs are proven, QNEC in particular. The proof of QNEC given in~\cite{Balakrishnan:2017bjg} is valid for all relativistic QFTs in Minkowski spacetime. However, there are also proofs of QNEC on curved backgrounds,
    for a smaller set of QFTs,
    such as in~\cite{Bousso:2014sda, Fu:2017evt}. It would be advantageous, particularly with the application to semi-classical gravity in mind, to generalise our work to curved backgrounds. One obstacle to overcome is the transverse integration of the previous bullet point. Because the integration region of our bound in $d>2$ dimensions could detect global structures of a background, we might incur large curvature corrections or even topological obstructions to our bound. A more local expression of the form mentioned in the previous bullet would allow us to work perturbatively when smearing over regions smaller than the typical curvature scale.
    
    \item Our derivations take QNEC as one of their inputs. In the same way that, while ANEC can be proven by integrating QNEC along a complete null line, it can also be proven directly using causality~\cite{Hartman:2016lgu} and monotonicity of relative entropy~\cite{Faulkner:2016mzt}, so we expect that our \boundName[s] can be proven without taking QNEC as input, perhaps using the causality and monotonicity of relative entropy.
    
    \item It would be interesting to extend our work to QFTs coupled to gravity. One intriguing possibility is that gravity itself provides a natural regulator of negative energy densities. This is essentially the core idea of the SNEC \eqref{eq:snecdef} mentioned in the introduction of this paper. However, as mentioned there, collective evidence for or even proving such QEIs is a hard task because they require coupling to gravity as input. Existing proofs of QNEC, such as~\cite{Balakrishnan:2017bjg}, assume a fixed background geometry, but the original statement of QNEC came as a corollary of the QFC~\cite{Bousso:2015mna}, and the motivation of that paper was to study QFTs coupled to gravity. 
    It would be interesting to see whether QEIs for QFTs coupled to gravity can be proven directly from the QFC. We are actively working in this direction and will report results in a future publication \cite{gravRolphBounds}. 

\end{itemize}

\acknowledgments

We would like to thank Tom Faulkner, Ben Freivogel, Eleni Kontou, Diego Pardo Santos, Gabriele Pascuzzi, Aron Wall, and Sasha Zhiboedov
for useful discussions and to Ben Freivogel in particular for initial collaboration and comments on a draft of this paper. We acknowledge the May 2024 COST Action 22113 - ``Fundamental Challenges in Theoretical Physics'' Kick-off Meeting at the University of Padova where this collaboration was initiated
. JRF additionally acknowledges the hospitality of CERN, Columbia University, and the Kavli Institute of Physics and Mathematics of the Universe where portions of this work were completed. The research of JRF is partially supported by STFC consolidated grants ST/T000694/1 and ST/X000664/1, partially by Simons Foundation Award number 620869, and partially by FNRS MISU grant 40024018 ``Pushing horizons in Black hole Physics.'' The work of AR is supported by FWO-Vlaanderen project G012222N, the VUB Research Council through the Strategic Research Program High-Energy Physics, and FWO-Vlaanderen through a Senior Postdoctoral Fellowship 1223125N.

\appendix

\section{Basics of functional differentiation.}\label{sec:func_diff}

Suppose we have a functional $F[f]$. The functional derivatives are defined through
\bne\label{eq:defdef}
    \delta F[f] := F[f+\delta f] - F[f] = \int \dd x\,\frac{\delta F}{\delta f(x)} \delta f(x)+ \frac{1}{2} \int \int \dd x\,\dd y\,\frac{\delta^2 F}{\delta f(x) \delta f(y)}\delta f (x) \delta f(y) + \dots
\ene

\textbf{Simple example 1:} $F[f] = \int \dd x\,f(x)^2$. Then
\bne
    \frac{\delta F}{\delta f(x)} = 2 f(x), \quad \frac{\delta^2 F}{\delta f(x) \delta f(y)} = 2 \delta (x-y)~,
\ene
and the higher variational derivatives vanish. Note that the 2nd variational derivative has a delta function, like in~\eqref{eq:2dv}. 

\textbf{Simple example 2:} $F[f] = \int \dd x\,f(x)^3$. Then
\begin{align}
    &\frac{\delta F}{\delta f(x)} = 3 f(x)^2~, \quad \frac{\delta^2 F}{\delta f(x) \delta f(y)} = 6 f(x) \delta (x-y)~,\nonumber\\
    &\frac{\delta^3 F}{\delta f(x) \delta f(y)\delta f(z)} = 6 \delta (x-z) \delta (y-z) = 6 \delta (y-z) \delta(x-y)~.
\end{align}

\textbf{Simple example 3 (non-local):}
$F[f] = \int \dd x\,\dd y\, f(x+y) f(x-y)$. We get
\bne
    \frac{\delta F}{\delta f(x)} = \int \dd y\,(f(x-2y) + f(x+2y)), \quad \frac{\delta^2 F}{\delta f(x)\delta f(y)} = 1~.
\ene
Note that the 2nd variational derivative has no delta-function, and is non-zero for $x\neq y$. In~\eqref{eq:2dv}, it is the non-locality of the entropy functional that allows it to have off-diagonal terms.

\textbf{Delta-function variations:} Suppose that $\delta f(x) = \lambda \delta (x-x_0)$. Plugging this into~\eqref{eq:defdef} gives
\bne 
    \delta F = \lambda \frac{\delta F}{\delta f(x_0)} + \frac{\lambda^2}{2} \frac{\delta^2 F}{\delta f(x_0)^2} + \dots 
\ene
and so
\begin{equation}
    \frac{\dd F}{\dd\lambda}\bigg|_{\lambda = 0} = \frac{\delta F}{\delta f(x_0)},\quad \frac{\dd^2 F}{\dd\lambda^2}\bigg|_{\lambda=0} = \frac{\delta^2 F}{\delta f(x_0)^2}, \quad \dots 
\end{equation}
This is sometimes how the variational derivatives are defined. Note that to define the ordinary derivative $\frac{\dd^n F}{\dd\lambda^n}$ we had to specify the variation $\delta f$. The variational derivatives do not depend on what $\delta f$ is.

\textbf{Box variations:} Suppose that the variation is box shaped: it equals $\lambda$ in some small neighbourhood of a point: $\delta f(x) = \lambda \Theta (\frac{\epsilon}{2} - |x -x_0|)$. Then we get
\bne
    \delta F \sim \lambda \epsilon \frac{\delta F}{\delta f(x_0)} \qquad \epsilon \to 0~.
\ene
To correctly go from this to the ordinary derivative, note we have to take $\epsilon \to 0$ first before taking $\lambda \to 0$, i.e. if we define
\bne
    \frac{1}{\epsilon} \frac{\dd F}{\dd\lambda}\bigg|_{\lambda = 0} := \lim_{\lambda \to 0} \lim_{\epsilon \to 0} \frac{F[f+ \lambda \Theta (\frac{\epsilon}{2} - |x -x_0|)] - F[f]}{\epsilon\lambda}~,
\ene
then we have
\bne
    \frac{1}{\epsilon} \frac{\dd F}{\dd\lambda}\bigg|_{\lambda = 0} =  \frac{\delta F}{\delta f(x_0)}~.
\ene
$\epsilon$ is a coordinate width, and so becomes $\cA/\sqrt{h}$ for higher-dimensional curved manifolds, where $\cA$ is the proper area of the top of the infinitesimal box shape variation.
This then connects to the notation used in~\cite{Bousso:2015mna}.

\section{Testing bounds with Ba\~nados states} \label{sec:testing_bounds}
We will test some of our inequalities against states in holographic 2d CFTs dual to Ba\~nados geometries~\cite{Sheikh-Jabbari:2016znt}. In these states, both the null energy density and the entanglement entropy are known.

For Ba\~nados states,
\bne 
    2\pi \langle T_{vv} (v) \rangle =  \frac{c}{6} L_v (v) 
\ene
and similarly for $v \to u$. 
$L_v$ is a diffeomorphism $v \mapsto L_v (v)$. 
We need the entanglement entropies of intervals in Ba\~nados states, and these are known for arbitrary $L_v$, but, for simplicity, let us restrict to constant non-negative $L_v$, which are dual to BTZ black holes. 
The entropy of an interval $(0,0)$ to $(u,v)$ splits into left and right-moving pieces
\bne 
    S (u,v) = S_u (u) + S_v (v) ~,
\ene
with, for the BTZ state, 
\bne 
    S_v = \frac{c}{6} \log \left ( \frac{\sinh \sqrt{L_v} v}{\sqrt{L_v} \epsilon}\right )~.
\ene
\textbf{2d QNEC.} From this, we have
\bne 
    \del_v^2 S + \frac{6}{c} (\del_v S)^2 = \frac{c}{6} L_v~,
\ene
which shows that the BTZ state saturates the 2d QNEC~\eqref{eq:2dQNE}.
\bne 
    2\pi T_{vv} \geq \del_v^2 S + \frac{6}{c} (\del_v S)^2~. 
\ene
As mentioned in Sec.~\ref{sec:2dSNEC}, \textit{all} Ba\~nados states saturate the 2d QNEC~\cite{Khandker:2018xls, Ecker:2019ocp}; what we have done here is verify that for constant $L_v$. 

\textbf{Bound on $\bm{\del_v S(v)}$.} Now we test the bound~\eqref{eq:upb} on $\del_v S(v)$,
When $\langle T_{vv}(v') \rangle$ is constant, as it is for the BTZ black hole, we can determine the optimising $\zeta$ which tightens the bound as much as possible. This is 
\bne 
    \zeta (v) = v + \sgn(v) \sqrt{\frac{3}{L_v}}~.
\ene
For a constant null energy, using the optimal $\zeta$,~\eqref{eq:upb} reduces to
\bne 
    \frac{c}{6} \sqrt{L_v} \tanh(\sqrt{L_v} v) \leq \frac{c}{3} \sqrt{ \frac{L_v}{3}}~,
\ene
which is indeed true.

\textbf{\boundName.} 
The \boundName we will check is~\eqref{eq:not_even_my_final_form}. The right-hand side vanishes when $v - \zeta(v)$ is constant, as we have. The left-hand side is non-negative, and so~\eqref{eq:not_even_my_final_form} is satisfied.

For the constant $L_v$ Ba\~nados states considered here, we have shown that they saturate the 2d QNEC, but not the 2d \boundName~\eqref{eq:not_even_my_final_form} we derived starting from integrating 2d QNEC. This is because the inequalities we used at intermediate steps in the derivation, such as strong subadditivity, are not saturated. This illustrates that in deriving the 2d \boundName[s], we have gained state-independence at the cost of weakening the bound on the stress tensor with respect to 2d QNEC.

\section{2d \boundName for non-invertible \texorpdfstring{\boldmath$\zeta$}{\zeta}} 
\label{sec:non_invertible_zeta}
Here we will extend the 2d \boundName[s] of Sec.~\ref{sect:Rolphbound} to invertible $v_B (v)$ and $v_C (v)$ but non-invertible $\zeta$. Defined in~\eqref{eq:zetde}, $\zeta$ is only invertible if $v_B (0) = v_C (0) = 0$. If $v_B(0) \neq 0$ or $v_C(0) \neq 0$, then $\zeta$ is injective but not surjective, and $\zeta^{-1}$ is only defined on the image of $\zeta$, which is $v \geq v_C (0) \geq 0$ and $v \leq v_B (0) \leq 0$. 

We start from~\eqref{eq:longq}, written as
\bne \begin{split} 
    2\pi \int_{-\infty}^{\infty} \dd v\, g(v) T_{vv}(v) &\geq \int_{-\infty}^0 \dd v\, g'(v) \left (-2\pi \int^v_{v_B(v)} \dd v' \left( \frac{v'-\zeta(v)}{v - \zeta(v)}\right)^2\langle T_{vv} (v') \rangle - \frac{\cUV}{6}\frac{1}{v- \zeta(v)} \right) \\
    & - \int_0^\infty \dd v\, g'(v) \left( -2\pi \int_v^{v_C (v)} \dd v' \left( \frac{v'-\zeta(v)}{v - \zeta(v)}\right)^2\langle T_{vv} (v') \rangle - \frac{\cUV}{6}\frac{1}{\zeta(v) - v} \right) .\\ \label{eq:inmrs}
\end{split} \ene

Next, we will switch the order of integration, using the relations%
\footnote{Note that this uses our assumption that $v_B$ and $v_C$ are invertible, i.e. their inverses are defined for the whole real line.}
\bne \begin{split} 
    \int_{-\infty}^0 \dd v \int_{v_B (v)}^v \dd v' &= \int_{-\infty}^0 \dd v' \int_{v'}^{\min (v_B^{-1}(v'),0)}\dd v ~.
\end{split} \ene
and 
\bne\begin{split} 
\int_0^\infty \dd v \int_v^{v_C (v)} \dd v' &= \int_0^\infty \dd v' \int^{v'}_{\max (0,v_C^{-1}(v'))}\dd v~.
\end{split}\ene
Then~\eqref{eq:inmrs} becomes
\bne 2\pi \int_{-\infty}^{\infty} dv (g(v)+h(v)) T_{vv}(v) \geq  -\frac{c}{6} \int_{-\infty}^\infty dv  \frac{g'(v)}{v-\zeta(v)}  
\label{eq:2dSN2}\ene
where
\bne h(v) = \begin{cases} \quad \int_{v}^{\min(v_B^{-1}(v),0)} dv' g'(v') \left (\frac{v - \zeta(v')}{v'-\zeta (v')}  \right)^2 &v<0 \\
-\int^{v}_{\max(0,v_C^{-1}(v))}dv' g'(v') \left (\frac{v - \zeta(v')}{v'-\zeta (v')} \right)^2 &0< v.
\label{eq:hv2de}
\end{cases}\ene
Alternatively, if we define the pseudoinverse
\beq
    \overline{\zeta^{-1}}(v)
    :=\begin{cases}   v_B^{-1}(v) &v\in(-\infty,v_B(0)] \\
    0 & v\in(-v_B(0),v_C(0))\\
    v_C^{-1}(v)&v\in[v_C(0),\infty)
\end{cases}
\eeq
then~\eqref{eq:hv2de} can also be written as
\beq
    h(v)=-\int_{\overline{\zeta^{-1}}(v')}^v\dd v'\,g'(v')\left(\frac{v-\zeta(v')}{v'-\zeta(v')}\right)^2~.
    \label{eq:hdef2}
\eeq

Eq.~\eqref{eq:2dSN2}, with $h$ given in~\eqref{eq:hv2de} and~\eqref{eq:hdef2}, is the generalisation of the 2d \boundName~\eqref{eq:not_even_my_final_form}, having dropped the assumption that $\zeta$ is invertible, though still assuming that $v_B$ and $v_C$ are invertible. As a consistency check, note that when $\zeta$ \textit{is} invertible (i.e. $v_B(0) = v_C(0) =0$), then $\min (v_B^{-1}(v'),0) = v_B^{-1}(v')$ and $\max (0,v_C^{-1}(v')) = v_C^{-1}(v') $, and~\eqref{eq:hv2de} matches~\eqref{eq:hdefn}.

\paragraph{Example.} Now we consider an example for which $\zeta$ is non-invertible: $v_B(v) = v-\delta$ and $v_C (v) = v+\delta$, with $\delta >0$. This is the optimal choice when $T_{vv}$ is positive and constant, see Eq.~\eqref{eq:const_T}. For this choice,
\bne h(v) = \begin{cases}  
I(v,0) & |v|<\delta \\
I(v,v-\delta \sgn(v)) &\delta < |v|
\label{eq:hvdel}
\end{cases}\ene
where
\bne I (v_1,v_2) := \delta^{-2} \int_{v_1}^{v_2} dv' g'(v') \left (v -v' - \delta\sgn(v_1)   \right)^2 
. \ene
From its definition, this $h(v)$ is non-negative, even, and continuous at $|v| =  \delta$, with $h(\delta) = I (\delta, 0)$. Also, $h(0) = 0$.

\begin{figure}
    \centering
    \includegraphics[width=0.8\linewidth]{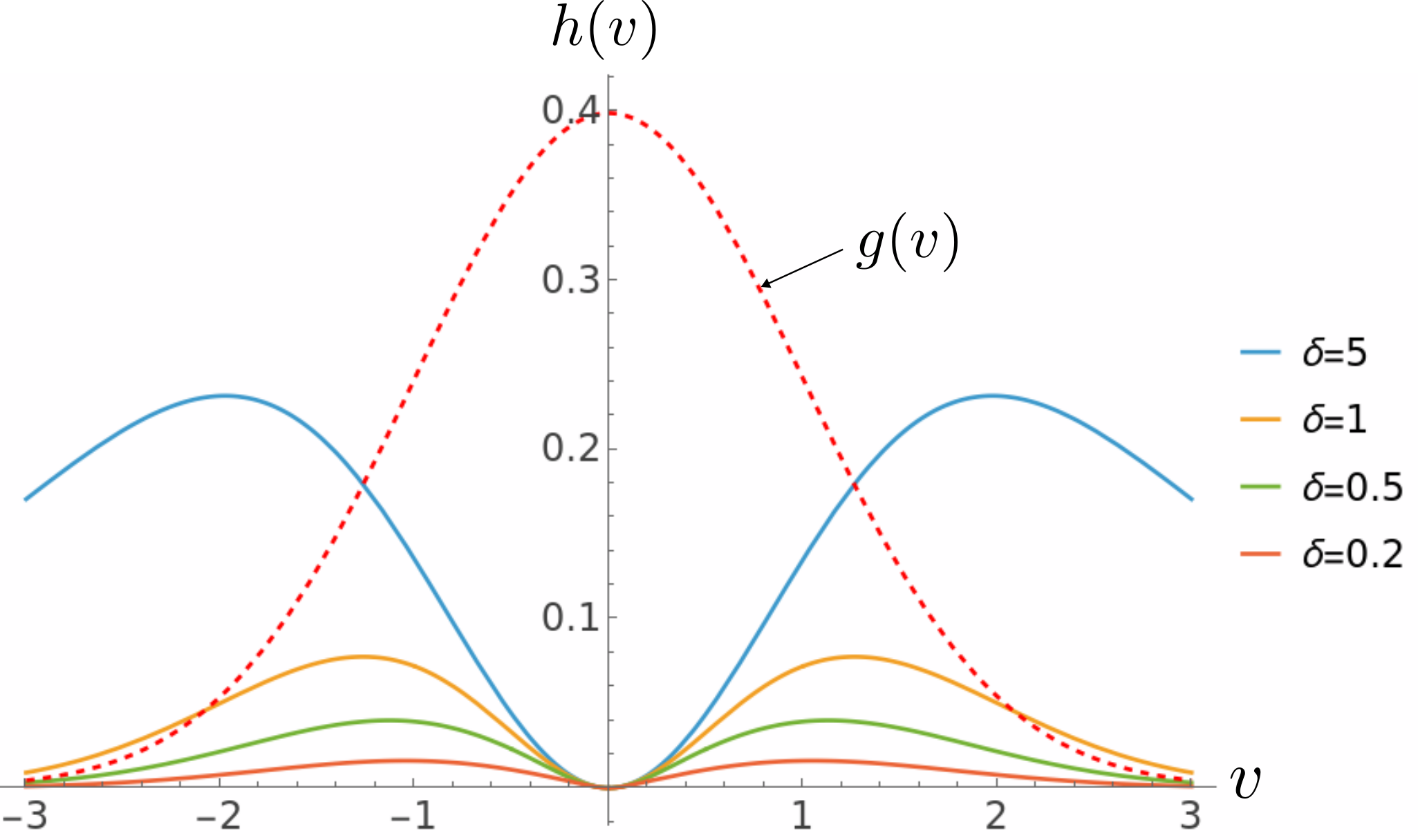}
    \caption{A plot of~\eqref{eq:hvdel}, which is $h(v)$ for the choice of $v_B(v) = v-\delta$ and $v_C(v) = v+\delta$. We take $g$ to be a normalised Gaussian.}
    \label{fig:hvplot}
\end{figure}

If we take the $\delta \to 0$ limit of $h$, we find
\bne h(v) = \begin{cases} 
\frac{v^2}{12\delta^2}(-g''(0)) (v^2 -4 |v|\delta +6\delta^2) + \cO(\delta^3) \qquad &|v| < \delta \\
-\frac{1}{3} \delta \sgn (v) g'(v) + \frac{1}{12} \delta^2 g''(v) + \cO (\delta^3) , \qquad  &\delta < |v| \end{cases} \ene
As promised, this $h(v)$ is non-negative, even, and continuous at $|v| = \delta$, with $h(\delta) = \frac{1}{4}\delta^2 (-g''(0)) + \cO(\delta^3)$. It is also $\cO (\delta)$, so,
for small $\delta$, the 2d \boundName~\eqref{eq:2dSN2} becomes
\bne 2\pi \int_{-\infty}^\infty dv g(v) \langle T_{vv}(v) \rangle + O(\delta) \geq -\frac{c}{3} \frac{g(0)}{\delta} .\ene
Note that this bound is valid for any ($\delta$-independent) $g$. This bound becomes trivial in the limit $\delta \to 0$, consistent with the case of $\zeta (v) =v$ discussed in Sec.~\ref{sec:SNEC_bounds}, but the bound is non-trivial for finite $\delta$.

Let us give an example for finite $\delta$. When $g(v)$ is a Gaussian
\bne g(v)= \frac{1}{\sqrt{2\pi}\sigma} e^{-v^2/2\sigma^2}\,, \ene
then $h$ is (defining $r=|v|$)
\bne
h(v) = \frac{1}{\delta^2}\begin{cases} 
 2\sigma^2(g(0)-g(r))-\delta^2 g(r) + 2(\delta-r) \Phi(r) + g(0)(\delta-r)^2, & r <\delta \\
 2\sigma^2(g(r-\delta)-  g(r))- \delta^2 g(r) - 2(r-\delta) (\Phi(r) - \Phi(r-\delta)), & \delta < r
\end{cases} 
\label{eq:hvhva}
\ene
where
\bne  \Phi(v) := \int_{-\infty}^v g(v')dv' = \frac{1}{2}\left(1+\erf\left(\frac{v}{\sqrt{2}\sigma} \right)\right). \ene
See Fig.~\ref{fig:hvplot} for a plot of~\eqref{eq:hvhva} for different values of $\delta$.

\bibliography{EnergyBounds}

@article{gravRolphBounds,
    author = "Fliss, Jackson R. and Rolph, Andrew",
    journal = "to appear."
}

@article{Fliss:2024dxe,
    author = "Fliss, Jackson R. and Freivogel, Ben and Kontou, Eleni-Alexandra and Santos, Diego Pardo",
    title = "{How negative can null energy be in large N CFTs?}",
    eprint = "2412.10618",
    archivePrefix = "arXiv",
    primaryClass = "hep-th",
    month = "12",
    year = "2024"
}

@article{Fewster:2007ec,
    author = "Fewster, Christopher J. and Osterbrink, Lutz W.",
    title = "{Quantum Energy Inequalities for the Non-Minimally Coupled Scalar Field}",
    eprint = "0708.2450",
    archivePrefix = "arXiv",
    primaryClass = "gr-qc",
    doi = "10.1088/1751-8113/41/2/025402",
    journal = "J. Phys. A",
    volume = "41",
    pages = "025402",
    year = "2008"
}

@article{Fewster:2021mmz,
    author = "Fewster, Christopher J. and Kontou, Eleni-Alexandra",
    title = "{A semiclassical singularity theorem}",
    eprint = "2108.12668",
    archivePrefix = "arXiv",
    primaryClass = "gr-qc",
    doi = "10.1088/1361-6382/ac566b",
    journal = "Class. Quant. Grav.",
    volume = "39",
    number = "7",
    pages = "075028",
    year = "2022"
}

@article{Fewster:2019bjg,
    author = "Fewster, Christopher J. and Kontou, Eleni-Alexandra",
    title = "{A new derivation of singularity theorems with weakened energy hypotheses}",
    eprint = "1907.13604",
    archivePrefix = "arXiv",
    primaryClass = "gr-qc",
    doi = "10.1088/1361-6382/ab685b",
    journal = "Class. Quant. Grav.",
    volume = "37",
    number = "6",
    pages = "065010",
    year = "2020"
}

@article{Brown:2018hym,
    author = "Brown, Peter J. and Fewster, Christopher J. and Kontou, Eleni-Alexandra",
    title = "{A singularity theorem for Einstein{\textendash}Klein{\textendash}Gordon theory}",
    eprint = "1803.11094",
    archivePrefix = "arXiv",
    primaryClass = "gr-qc",
    doi = "10.1007/s10714-018-2446-5",
    journal = "Gen. Rel. Grav.",
    volume = "50",
    number = "10",
    pages = "121",
    year = "2018"
}

@article{Hawking:1966sx,
    author = "Hawking, Stephen",
    title = "{The Occurrence of singularities in cosmology}",
    doi = "10.1098/rspa.1966.0221",
    journal = "Proc. Roy. Soc. Lond. A",
    volume = "294",
    pages = "511--521",
    year = "1966"
}

@article{Hawking:1966jv,
    author = "Hawking, Stephen",
    title = "{The Occurrence of singularities in cosmology. II}",
    doi = "10.1098/rspa.1966.0255",
    journal = "Proc. Roy. Soc. Lond. A",
    volume = "295",
    pages = "490--493",
    year = "1966"
}

@article{Hawking:1967ju,
    author = "Hawking, Stephen",
    title = "{The occurrence of singularities in cosmology. III. Causality and singularities}",
    doi = "10.1098/rspa.1967.0164",
    journal = "Proc. Roy. Soc. Lond. A",
    volume = "300",
    pages = "187--201",
    year = "1967"
}

@inproceedings{Faulkner:2022mlp,
    author = "Faulkner, Thomas and Hartman, Thomas and Headrick, Matthew and Rangamani, Mukund and Swingle, Brian",
    title = "{Snowmass white paper: Quantum information in quantum field theory and quantum gravity}",
    booktitle = "{Snowmass 2021}",
    eprint = "2203.07117",
    archivePrefix = "arXiv",
    primaryClass = "hep-th",
    reportNumber = "BRX-TH-6703",
    month = "3",
    year = "2022"
}

@article{Bekenstein:1974ax,
    author = "Bekenstein, Jacob D.",
    title = "{Generalized second law of thermodynamics in black hole physics}",
    doi = "10.1103/PhysRevD.9.3292",
    journal = "Phys. Rev. D",
    volume = "9",
    pages = "3292--3300",
    year = "1974"
}

@article{Bousso:1999xy,
    author = "Bousso, Raphael",
    title = "{A Covariant entropy conjecture}",
    eprint = "hep-th/9905177",
    archivePrefix = "arXiv",
    reportNumber = "SU-ITP-99-23",
    doi = "10.1088/1126-6708/1999/07/004",
    journal = "JHEP",
    volume = "07",
    pages = "004",
    year = "1999"
}

@article{Bousso:2022tdb,
    author = "Bousso, Raphael and Shahbazi-Moghaddam, Arvin",
    title = "{Quantum singularities}",
    eprint = "2206.07001",
    archivePrefix = "arXiv",
    primaryClass = "hep-th",
    doi = "10.1103/PhysRevD.107.066002",
    journal = "Phys. Rev. D",
    volume = "107",
    number = "6",
    pages = "066002",
    year = "2023"
}

@article{Bousso:2022cun,
    author = "Bousso, Raphael and Shahbazi-Moghaddam, Arvin",
    title = "{Singularities from Entropy}",
    eprint = "2201.11132",
    archivePrefix = "arXiv",
    primaryClass = "hep-th",
    doi = "10.1103/PhysRevLett.128.231301",
    journal = "Phys. Rev. Lett.",
    volume = "128",
    number = "23",
    pages = "231301",
    year = "2022"
}

@article{Bousso:2025xyc,
    author = "Bousso, Raphael",
    title = "{Robust Singularity Theorem}",
    eprint = "2501.17910",
    archivePrefix = "arXiv",
    primaryClass = "hep-th",
    doi = "10.1103/6f9b-3jmx",
    journal = "Phys. Rev. Lett.",
    volume = "135",
    number = "1",
    pages = "011501",
    year = "2025"
}

@article{Bousso:2015mna,
    author = "Bousso, Raphael and Fisher, Zachary and Leichenauer, Stefan and Wall, Aron C.",
    title = "{Quantum focusing conjecture}",
    eprint = "1506.02669",
    archivePrefix = "arXiv",
    primaryClass = "hep-th",
    doi = "10.1103/PhysRevD.93.064044",
    journal = "Phys. Rev. D",
    volume = "93",
    number = "6",
    pages = "064044",
    year = "2016"
}

@article{Wall:2010jtc,
    author = "Wall, Aron C.",
    title = "{The Generalized Second Law implies a Quantum Singularity Theorem}",
    eprint = "1010.5513",
    archivePrefix = "arXiv",
    primaryClass = "gr-qc",
    doi = "10.1088/0264-9381/30/19/199501",
    journal = "Class. Quant. Grav.",
    volume = "30",
    pages = "165003",
    year = "2013",
    note = "[Erratum: Class.Quant.Grav. 30, 199501 (2013)]"
}

@article{Reeh:1961ujh,
    author = "Reeh, H. and Schlieder, S.",
    title = {{Bemerkungen zur unit{\"a}r{\"a}quivalenz von lorentzinvarianten feldern}},
    doi = "10.1007/BF02787889",
    journal = "Nuovo Cim.",
    volume = "22",
    number = "5",
    pages = "1051--1068",
    year = "1961"
}

@article{Zamolodchikov:1986gt,
    author = "Zamolodchikov, A. B.",
    title = "{Irreversibility of the Flux of the Renormalization Group in a 2D Field Theory}",
    journal = "JETP Lett.",
    volume = "43",
    pages = "730--732",
    year = "1986"
}

@article{Burkardt:1995ct,
    author = "Burkardt, Matthias",
    title = "{Light front quantization}",
    eprint = "hep-ph/9505259",
    archivePrefix = "arXiv",
    doi = "10.1007/0-306-47067-5_1",
    journal = "Adv. Nucl. Phys.",
    volume = "23",
    pages = "1--74",
    year = "1996"
}

@article{jrf_modave,
    author = "Fliss, Jackson R.",
    title = "{Modave lectures on quantum energy inequalities.}",
    journal = "to appear."
}

@article{Osborn:1993cr,
    author = "Osborn, H. and Petkou, A. C.",
    title = "{Implications of conformal invariance in field theories for general dimensions}",
    eprint = "hep-th/9307010",
    archivePrefix = "arXiv",
    reportNumber = "DAMTP-93-31",
    doi = "10.1006/aphy.1994.1045",
    journal = "Annals Phys.",
    volume = "231",
    pages = "311--362",
    year = "1994"
}

@article{Buchel:2009sk,
    author = "Buchel, Alex and Escobedo, Jorge and Myers, Robert C. and Paulos, Miguel F. and Sinha, Aninda and Smolkin, Michael",
    title = "{Holographic GB gravity in arbitrary dimensions}",
    eprint = "0911.4257",
    archivePrefix = "arXiv",
    primaryClass = "hep-th",
    reportNumber = "UWO-TH-09-16",
    doi = "10.1007/JHEP03(2010)111",
    journal = "JHEP",
    volume = "03",
    pages = "111",
    year = "2010"
}

@article{Ryu:2006bv,
    author = "Ryu, Shinsei and Takayanagi, Tadashi",
    title = "{Holographic derivation of entanglement entropy from AdS/CFT}",
    eprint = "hep-th/0603001",
    archivePrefix = "arXiv",
    reportNumber = "NSF-KITP-06-11, NSF-KITP-06-11",
    doi = "10.1103/PhysRevLett.96.181602",
    journal = "Phys. Rev. Lett.",
    volume = "96",
    pages = "181602",
    year = "2006"
}

@article{Casini:2017roe,
    author = "Casini, Horacio and Teste, Eduardo and Torroba, Gonzalo",
    title = "{Modular Hamiltonians on the null plane and the Markov property of the vacuum state}",
    eprint = "1703.10656",
    archivePrefix = "arXiv",
    primaryClass = "hep-th",
    doi = "10.1088/1751-8121/aa7eaa",
    journal = "J. Phys. A",
    volume = "50",
    number = "36",
    pages = "364001",
    year = "2017"
}

@article{Bousso:2014uxa,
    author = "Bousso, Raphael and Casini, Horacio and Fisher, Zachary and Maldacena, Juan",
    title = "{Entropy on a null surface for interacting quantum field theories and the Bousso bound}",
    eprint = "1406.4545",
    archivePrefix = "arXiv",
    primaryClass = "hep-th",
    doi = "10.1103/PhysRevD.91.084030",
    journal = "Phys. Rev. D",
    volume = "91",
    number = "8",
    pages = "084030",
    year = "2015"
}

@article{Faulkner:2015csl,
    author = "Faulkner, Thomas and Leigh, Robert G. and Parrikar, Onkar",
    title = "{Shape Dependence of Entanglement Entropy in Conformal Field Theories}",
    eprint = "1511.05179",
    archivePrefix = "arXiv",
    primaryClass = "hep-th",
    doi = "10.1007/JHEP04(2016)088",
    journal = "JHEP",
    volume = "04",
    pages = "088",
    year = "2016"
}

@article{Sheikh-Jabbari:2016znt,
    author = "Sheikh-Jabbari, M. M. and Yavartanoo, H.",
    title = "{Excitation entanglement entropy in two dimensional conformal field theories}",
    eprint = "1605.00341",
    archivePrefix = "arXiv",
    primaryClass = "hep-th",
    reportNumber = "NO.12, 126006",
    doi = "10.1103/PhysRevD.94.126006",
    journal = "Phys. Rev. D",
    volume = "94",
    number = "12",
    pages = "126006",
    year = "2016"
}

@article{Ecker:2019ocp,
    author = "Ecker, Christian and Grumiller, Daniel and van der Schee, Wilke and Sheikh-Jabbari, M. M. and Stanzer, Philipp",
    title = "{Quantum Null Energy Condition and its (non)saturation in 2d CFTs}",
    eprint = "1901.04499",
    archivePrefix = "arXiv",
    primaryClass = "hep-th",
    reportNumber = "TUW-18-07, IPM/P-2018/063",
    doi = "10.21468/SciPostPhys.6.3.036",
    journal = "SciPost Phys.",
    volume = "6",
    number = "3",
    pages = "036",
    year = "2019"
}

@article{Casini:2022rlv,
    author = "Casini, Horacio and Huerta, Marina",
    title = "{Lectures on entanglement in quantum field theory}",
    eprint = "2201.13310",
    archivePrefix = "arXiv",
    primaryClass = "hep-th",
    doi = "10.22323/1.403.0002",
    journal = "PoS",
    volume = "TASI2021",
    pages = "002",
    year = "2023"
}

@article{Wall:2011kb,
    author = "Wall, Aron C.",
    title = "{Testing the Generalized Second Law in 1+1 dimensional Conformal Vacua: An Argument for the Causal Horizon}",
    eprint = "1105.3520",
    archivePrefix = "arXiv",
    primaryClass = "gr-qc",
    doi = "10.1103/PhysRevD.85.024015",
    journal = "Phys. Rev. D",
    volume = "85",
    pages = "024015",
    year = "2012"
}

@article{Bousso:2015wca,
    author = "Bousso, Raphael and Fisher, Zachary and Koeller, Jason and Leichenauer, Stefan and Wall, Aron C.",
    title = "{Proof of the Quantum Null Energy Condition}",
    eprint = "1509.02542",
    archivePrefix = "arXiv",
    primaryClass = "hep-th",
    doi = "10.1103/PhysRevD.93.024017",
    journal = "Phys. Rev. D",
    volume = "93",
    number = "2",
    pages = "024017",
    year = "2016"
}

@article{Balakrishnan:2017bjg,
    author = "Balakrishnan, Srivatsan and Faulkner, Thomas and Khandker, Zuhair U. and Wang, Huajia",
    title = "{A General Proof of the Quantum Null Energy Condition}",
    eprint = "1706.09432",
    archivePrefix = "arXiv",
    primaryClass = "hep-th",
    doi = "10.1007/JHEP09(2019)020",
    journal = "JHEP",
    volume = "09",
    pages = "020",
    year = "2019"
}

@article{Ceyhan:2018zfg,
    author = "Ceyhan, Fikret and Faulkner, Thomas",
    title = "{Recovering the QNEC from the ANEC}",
    eprint = "1812.04683",
    archivePrefix = "arXiv",
    primaryClass = "hep-th",
    doi = "10.1007/s00220-020-03751-y",
    journal = "Commun. Math. Phys.",
    volume = "377",
    number = "2",
    pages = "999--1045",
    year = "2020"
}

@article{Hartman:2016lgu,
    author = "Hartman, Thomas and Kundu, Sandipan and Tajdini, Amirhossein",
    title = "{Averaged Null Energy Condition from Causality}",
    eprint = "1610.05308",
    archivePrefix = "arXiv",
    primaryClass = "hep-th",
    doi = "10.1007/JHEP07(2017)066",
    journal = "JHEP",
    volume = "07",
    pages = "066",
    year = "2017"
}

@article{Balakrishnan:2019gxl,
    author = "Balakrishnan, Srivatsan and Chandrasekaran, Venkatesa and Faulkner, Thomas and Levine, Adam and Shahbazi-Moghaddam, Arvin",
    title = "{Entropy variations and light ray operators from replica defects}",
    eprint = "1906.08274",
    archivePrefix = "arXiv",
    primaryClass = "hep-th",
    doi = "10.1007/JHEP09(2022)217",
    journal = "JHEP",
    volume = "09",
    pages = "217",
    year = "2022"
}

@article{Leichenauer:2018obf,
    author = "Leichenauer, Stefan and Levine, Adam and Shahbazi-Moghaddam, Arvin",
    title = "{Energy density from second shape variations of the von Neumann entropy}",
    eprint = "1802.02584",
    archivePrefix = "arXiv",
    primaryClass = "hep-th",
    doi = "10.1103/PhysRevD.98.086013",
    journal = "Phys. Rev. D",
    volume = "98",
    number = "8",
    pages = "086013",
    year = "2018"
}

@article{Koeller:2015qmn,
    author = "Koeller, Jason and Leichenauer, Stefan",
    title = "{Holographic Proof of the Quantum Null Energy Condition}",
    eprint = "1512.06109",
    archivePrefix = "arXiv",
    primaryClass = "hep-th",
    doi = "10.1103/PhysRevD.94.024026",
    journal = "Phys. Rev. D",
    volume = "94",
    number = "2",
    pages = "024026",
    year = "2016"
}

@article{Leichenauer:2018tnq,
    author = "Leichenauer, Stefan and Levine, Adam",
    title = "{Upper and Lower Bounds on the Integrated Null Energy in Gravity}",
    eprint = "1808.09970",
    archivePrefix = "arXiv",
    primaryClass = "hep-th",
    doi = "10.1007/JHEP01(2019)133",
    journal = "JHEP",
    volume = "01",
    pages = "133",
    year = "2019"
}

@article{Fliss:2023rzi,
    author = "Fliss, Jackson R. and Freivogel, Ben and Kontou, Eleni-Alexandra and Santos, Diego Pardo",
    title = "{Non-minimal coupling, negative null energy, and effective field theory}",
    eprint = "2309.10848",
    archivePrefix = "arXiv",
    primaryClass = "hep-th",
    doi = "10.21468/SciPostPhys.16.5.119",
    journal = "SciPost Phys.",
    volume = "16",
    number = "5",
    pages = "119",
    year = "2024"
}

@article{Fliss:2021gdz,
    author = "Fliss, Jackson R. and Freivogel, Ben",
    title = "{Semi-local Bounds on Null Energy in QFT}",
    eprint = "2108.06068",
    archivePrefix = "arXiv",
    primaryClass = "hep-th",
    doi = "10.21468/SciPostPhys.12.3.084",
    journal = "SciPost Phys.",
    volume = "12",
    number = "3",
    pages = "084",
    year = "2022"
}

@article{Freivogel:2018gxj,
    author = "Freivogel, Ben and Krommydas, Dimitrios",
    title = "{The Smeared Null Energy Condition}",
    eprint = "1807.03808",
    archivePrefix = "arXiv",
    primaryClass = "hep-th",
    doi = "10.1007/JHEP12(2018)067",
    journal = "JHEP",
    volume = "12",
    pages = "067",
    year = "2018"
}

@article{Wall:2011hj,
    author = "Wall, Aron C.",
    title = "{A proof of the generalized second law for rapidly changing fields and arbitrary horizon slices}",
    eprint = "1105.3445",
    archivePrefix = "arXiv",
    primaryClass = "gr-qc",
    doi = "10.1103/PhysRevD.85.104049",
    journal = "Phys. Rev. D",
    volume = "85",
    pages = "104049",
    year = "2012",
    note = "[Erratum: Phys.Rev.D 87, 069904 (2013)]"
}

@article{Bousso:2014sda,
    author = "Bousso, Raphael and Casini, Horacio and Fisher, Zachary and Maldacena, Juan",
    title = "{Proof of a Quantum Bousso Bound}",
    eprint = "1404.5635",
    archivePrefix = "arXiv",
    primaryClass = "hep-th",
    doi = "10.1103/PhysRevD.90.044002",
    journal = "Phys. Rev. D",
    volume = "90",
    number = "4",
    pages = "044002",
    year = "2014"
}

@article{Fewster:2002ne,
    author = "Fewster, Christopher J. and Roman, Thomas A.",
    title = "{Null energy conditions in quantum field theory}",
    eprint = "gr-qc/0209036",
    archivePrefix = "arXiv",
    doi = "10.1103/PhysRevD.67.044003",
    journal = "Phys. Rev. D",
    volume = "67",
    pages = "044003",
    year = "2003",
    note = "[Erratum: Phys.Rev.D 80, 069903 (2009)]"
}

@article{Faulkner:2016mzt,
    author = "Faulkner, Thomas and Leigh, Robert G. and Parrikar, Onkar and Wang, Huajia",
    title = "{Modular Hamiltonians for Deformed Half-Spaces and the Averaged Null Energy Condition}",
    eprint = "1605.08072",
    archivePrefix = "arXiv",
    primaryClass = "hep-th",
    doi = "10.1007/JHEP09(2016)038",
    journal = "JHEP",
    volume = "09",
    pages = "038",
    year = "2016"
}

@article{Fewster:2004nj,
    author = "Fewster, Christopher J. and Hollands, Stefan",
    title = "{Quantum energy inequalities in two-dimensional conformal field theory}",
    eprint = "math-ph/0412028",
    archivePrefix = "arXiv",
    doi = "10.1142/S0129055X05002406",
    journal = "Rev. Math. Phys.",
    volume = "17",
    pages = "577",
    year = "2005"
}

@article{Fewster:2012yh,
    author = "Fewster, Christopher J.",
    title = "{Lectures on quantum energy inequalities}",
    eprint = "1208.5399",
    archivePrefix = "arXiv",
    primaryClass = "gr-qc",
    month = "8",
    year = "2012"
}

@article{Headrick:2010zt,
    author = "Headrick, Matthew",
    title = "{Entanglement Renyi entropies in holographic theories}",
    eprint = "1006.0047",
    archivePrefix = "arXiv",
    primaryClass = "hep-th",
    reportNumber = "BRX-TH-619",
    doi = "10.1103/PhysRevD.82.126010",
    journal = "Phys. Rev. D",
    volume = "82",
    pages = "126010",
    year = "2010"
}

@article{Holzhey:1994we,
    author = "Holzhey, Christoph and Larsen, Finn and Wilczek, Frank",
    title = "{Geometric and renormalized entropy in conformal field theory}",
    eprint = "hep-th/9403108",
    archivePrefix = "arXiv",
    reportNumber = "PUPT-1454, IASSNS-HEP-93-88",
    doi = "10.1016/0550-3213(94)90402-2",
    journal = "Nucl. Phys. B",
    volume = "424",
    pages = "443--467",
    year = "1994"
}

@article{Casini:2011kv,
    author = "Casini, Horacio and Huerta, Marina and Myers, Robert C.",
    title = "{Towards a derivation of holographic entanglement entropy}",
    eprint = "1102.0440",
    archivePrefix = "arXiv",
    primaryClass = "hep-th",
    doi = "10.1007/JHEP05(2011)036",
    journal = "JHEP",
    volume = "05",
    pages = "036",
    year = "2011"
}

@article{Khandker:2018xls,
    author = "Khandker, Zuhair U. and Kundu, Sandipan and Li, Daliang",
    title = "{Bulk Matter and the Boundary Quantum Null Energy Condition}",
    eprint = "1803.03997",
    archivePrefix = "arXiv",
    primaryClass = "hep-th",
    doi = "10.1007/JHEP08(2018)162",
    journal = "JHEP",
    volume = "08",
    pages = "162",
    year = "2018"
}

@article{lieb1973proof,
  title={Proof of the strong subadditivity of quantum-mechanical entropy},
  author={Lieb, Elliott H and Ruskai, Mary Beth},
  journal={Les rencontres physiciens-math{\'e}maticiens de Strasbourg-RCP25},
  volume={19},
  pages={36--55},
  year={1973}
}

@article{epstein1965nonpositivity,
  title={Nonpositivity of the energy density in quantized field theories},
  author={Epstein, Henri and Glaser, Vladimir and Jaffe, Arthur},
  journal={Il Nuovo Cimento (1955-1965)},
  volume={36},
  number={3},
  pages={1016--1022},
  year={1965},
  publisher={Springer}
}

@article{Fu:2017evt,
    author = "Fu, Zicao and Koeller, Jason and Marolf, Donald",
    title = "{The Quantum Null Energy Condition in Curved Space}",
    eprint = "1706.01572",
    archivePrefix = "arXiv",
    primaryClass = "hep-th",
    doi = "10.1088/1361-6382/aa8f2c",
    journal = "Class. Quant. Grav.",
    volume = "34",
    number = "22",
    pages = "225012",
    year = "2017",
    note = "[Erratum: Class.Quant.Grav. 35, 049501 (2018)]"
}

@article{morris1988wormholes,
  title={Wormholes, time machines, and the weak energy condition},
  author={Morris, Michael S and Thorne, Kip S and Yurtsever, Ulvi},
  journal={Physical Review Letters},
  volume={61},
  number={13},
  pages={1446},
  year={1988},
  publisher={APS}
}

@article{Hochberg:1998ii,
    author = "Hochberg, David and Visser, Matt",
    title = "{The Null energy condition in dynamic wormholes}",
    eprint = "gr-qc/9802048",
    archivePrefix = "arXiv",
    reportNumber = "LAEFF-98-02",
    doi = "10.1103/PhysRevLett.81.746",
    journal = "Phys. Rev. Lett.",
    volume = "81",
    pages = "746--749",
    year = "1998"
}

@article{penrose1965gravitational,
  title={Gravitational collapse and space-time singularities},
  author={Penrose, Roger},
  journal={Physical Review Letters},
  volume={14},
  number={3},
  pages={57},
  year={1965},
  publisher={APS}
}

@article{Kontou:2015yha,
    author = "Kontou, Eleni-Alexandra and Olum, Ken D.",
    title = "{Proof of the averaged null energy condition in a classical curved spacetime using a null-projected quantum inequality}",
    eprint = "1507.00297",
    archivePrefix = "arXiv",
    primaryClass = "gr-qc",
    doi = "10.1103/PhysRevD.92.124009",
    journal = "Phys. Rev. D",
    volume = "92",
    pages = "124009",
    year = "2015"
}

@article{Wald:1991xn,
    author = "Wald, Robert M. and Yurtsever, U.",
    title = "{General proof of the averaged null energy condition for a massless scalar field in two-dimensional curved space-time}",
    doi = "10.1103/PhysRevD.44.403",
    journal = "Phys. Rev. D",
    volume = "44",
    pages = "403--416",
    year = "1991"
}

@article{Kelly:2014mra,
    author = "Kelly, William R. and Wall, Aron C.",
    title = "{Holographic proof of the averaged null energy condition}",
    eprint = "1408.3566",
    archivePrefix = "arXiv",
    primaryClass = "gr-qc",
    doi = "10.1103/PhysRevD.90.106003",
    journal = "Phys. Rev. D",
    volume = "90",
    number = "10",
    pages = "106003",
    year = "2014",
    note = "[Erratum: Phys.Rev.D 91, 069902 (2015)]"
}

@article{Gao:2000ga,
    author = "Gao, Sijie and Wald, Robert M.",
    title = "{Theorems on gravitational time delay and related issues}",
    eprint = "gr-qc/0007021",
    archivePrefix = "arXiv",
    doi = "10.1088/0264-9381/17/24/305",
    journal = "Class. Quant. Grav.",
    volume = "17",
    pages = "4999--5008",
    year = "2000"
}

@article{Rolph:2022csa,
    author = "Rolph, Andrew",
    title = "{Island mirages}",
    eprint = "2206.06144",
    archivePrefix = "arXiv",
    primaryClass = "hep-th",
    doi = "10.1007/JHEP08(2022)142",
    journal = "JHEP",
    volume = "08",
    pages = "142",
    year = "2022"
}

@article{Blanco:2017akw,
    author = "Blanco, David and Casini, Horacio and Leston, Mauricio and Rosso, Felipe",
    title = "{Modular energy inequalities from relative entropy}",
    eprint = "1711.04816",
    archivePrefix = "arXiv",
    primaryClass = "hep-th",
    doi = "10.1007/JHEP01(2018)154",
    journal = "JHEP",
    volume = "01",
    pages = "154",
    year = "2018"
}

@article{Blanco:2013lea,
    author = "Blanco, David D. and Casini, Horacio",
    title = "{Localization of Negative Energy and the Bekenstein Bound}",
    eprint = "1309.1121",
    archivePrefix = "arXiv",
    primaryClass = "hep-th",
    doi = "10.1103/PhysRevLett.111.221601",
    journal = "Phys. Rev. Lett.",
    volume = "111",
    number = "22",
    pages = "221601",
    year = "2013"
}

@article{Hollands:2025glm,
    author = "Hollands, Stefan and Longo, Roberto",
    title = "{A New Proof of the QNEC}",
    eprint = "2503.04651",
    archivePrefix = "arXiv",
    primaryClass = "hep-th",
    doi = "10.1007/s00220-025-05450-y",
    journal = "Commun. Math. Phys.",
    volume = "406",
    number = "11",
    pages = "269",
    year = "2025"
}

@misc{WittenTalk,
  author       = {Witten, Edward},
  title        = {Some Comments on Energy Inequalities},
  howpublished = {Talk at IAS Online Workshop on Qubits and Black Holes},
  year         = {2020},
  note         = {Recording available at www.youtube.com/watch?v=0Oh-Kmy-mx0.},
  url          = {www.youtube.com/watch?v=0Oh-Kmy-mx0}
}
\bibliographystyle{JHEP}
\end{document}